\documentclass[twocolumn,letterpaper,aps,prc,longbibliography,superscriptaddress,nofootinbib,floatfix]{revtex4-2}

\usepackage{graphicx}	

\usepackage{amsmath}	
\usepackage{multirow}
\usepackage{amssymb}
\usepackage{bm}

\usepackage{xspace}	
\newcommand{\pt}{\mbox{$p_{T}$}\xspace}
\newcommand{\raa}{\mbox{$R_{AA}$}\xspace}
\newcommand{\raace}{\mbox{$R_{AA}^{\ c{\rightarrow}e}$}\xspace}
\newcommand{\raabe}{\mbox{$R_{AA}^{\ b{\rightarrow}e}$}\xspace}

\newcommand{\npart}{\mbox{$N_{\rm part}$}\xspace}

\newcommand{\sqsn}{\mbox{$\sqrt{s_{_{NN}}}$}\xspace}
\newcommand{\snn}{\mbox{$\sqrt{s}$}\xspace}

\newcommand{\gevc}{\mbox{GeV/$c$}\xspace}
\newcommand{\gevcsq}{\mbox{GeV/$c^{2}$}\xspace}

\newcommand{\midy}{\mbox{$|y|<0.35$}\xspace}

\newcommand{\cbtoe}{\mbox{$c+b{\rightarrow}e$}\xspace}
\newcommand{\ctoe}{\mbox{$c{\rightarrow}e$}\xspace}
\newcommand{\btoe}{\mbox{$b{\rightarrow}e$}\xspace}
\newcommand{\dca}{\mbox{$\rm{DCA}$}\xspace}
\newcommand{\dcat}{\mbox{$\rm{DCA}_{T}$}\xspace}

\newcommand{\jps}{\mbox{J/$\psi$}\xspace}

\newcommand{\cherenkov}{\mbox{$\rm{\check{C}erenkov}$}\xspace}
\newcommand{\dep}{\mbox{\bf dep}\xspace}

\newcommand{\geant}{{\sc geant}-3\xspace}

\newcommand{\pp}{\mbox{$p$$+$$p$}\xspace}

\newcommand{\auau}{\mbox{Au$+$Au}\xspace}

\begin{document}

\title{Charm- and Bottom-Quark Production in Au$+$Au Collisions at 
$\sqrt{s_{_{NN}}}$ = 200 GeV}

\newcommand{\abilene}{Abilene Christian University, Abilene, Texas 79699, USA}
\newcommand{\augie}{Department of Physics, Augustana University, Sioux Falls, South Dakota 57197, USA}
\newcommand{\banaras}{Department of Physics, Banaras Hindu University, Varanasi 221005, India}
\newcommand{\barc}{Bhabha Atomic Research Centre, Bombay 400 085, India}
\newcommand{\baruch}{Baruch College, City University of New York, New York, New York, 10010 USA}
\newcommand{\bnlcoll}{Collider-Accelerator Department, Brookhaven National Laboratory, Upton, New York 11973-5000, USA}
\newcommand{\bnlphys}{Physics Department, Brookhaven National Laboratory, Upton, New York 11973-5000, USA}
\newcommand{\caucr}{University of California-Riverside, Riverside, California 92521, USA}
\newcommand{\charlesczech}{Charles University, Faculty of Mathematics and Physics, 180 00 Troja, Prague, Czech Republic}
\newcommand{\ciae}{Science and Technology on Nuclear Data Laboratory, China Institute of Atomic Energy, Beijing 102413, People's Republic of China}
\newcommand{\cns}{Center for Nuclear Study, Graduate School of Science, University of Tokyo, 7-3-1 Hongo, Bunkyo, Tokyo 113-0033, Japan}
\newcommand{\colorado}{University of Colorado, Boulder, Colorado 80309, USA}
\newcommand{\columbia}{Columbia University, New York, New York 10027 and Nevis Laboratories, Irvington, New York 10533, USA}
\newcommand{\czechtech}{Czech Technical University, Zikova 4, 166 36 Prague 6, Czech Republic}
\newcommand{\debrecen}{Debrecen University, H-4010 Debrecen, Egyetem t{\'e}r 1, Hungary}
\newcommand{\elte}{ELTE, E{\"o}tv{\"o}s Lor{\'a}nd University, H-1117 Budapest, P{\'a}zm{\'a}ny P.~s.~1/A, Hungary}
\newcommand{\ewha}{Ewha Womans University, Seoul 120-750, Korea}
\newcommand{\famu}{Florida A\&M University, Tallahassee, FL 32307, USA}
\newcommand{\fsu}{Florida State University, Tallahassee, Florida 32306, USA}
\newcommand{\gsu}{Georgia State University, Atlanta, Georgia 30303, USA}
\newcommand{\hiroshima}{Physics Program and International Institute for Sustainability with Knotted Chiral Meta Matter (SKCM2), Hiroshima University, Higashi-Hiroshima, Hiroshima 739-8526, Japan}
\newcommand{\howard}{Department of Physics and Astronomy, Howard University, Washington, DC 20059, USA}
\newcommand{\ihepprot}{IHEP Protvino, State Research Center of Russian Federation, Institute for High Energy Physics, Protvino, 142281, Russia}
\newcommand{\illuiuc}{University of Illinois at Urbana-Champaign, Urbana, Illinois 61801, USA}
\newcommand{\inrras}{Institute for Nuclear Research of the Russian Academy of Sciences, prospekt 60-letiya Oktyabrya 7a, Moscow 117312, Russia}
\newcommand{\instpasczech}{Institute of Physics, Academy of Sciences of the Czech Republic, Na Slovance 2, 182 21 Prague 8, Czech Republic}
\newcommand{\isu}{Iowa State University, Ames, Iowa 50011, USA}
\newcommand{\jaea}{Advanced Science Research Center, Japan Atomic Energy Agency, 2-4 Shirakata Shirane, Tokai-mura, Naka-gun, Ibaraki-ken 319-1195, Japan}
\newcommand{\jeonbuk}{Jeonbuk National University, Jeonju, 54896, Korea}
\newcommand{\jyvaskyla}{Helsinki Institute of Physics and University of Jyv{\"a}skyl{\"a}, P.O.Box 35, FI-40014 Jyv{\"a}skyl{\"a}, Finland}
\newcommand{\kek}{KEK, High Energy Accelerator Research Organization, Tsukuba, Ibaraki 305-0801, Japan}
\newcommand{\korea}{Korea University, Seoul 02841, Korea}
\newcommand{\kurchatov}{National Research Center ``Kurchatov Institute", Moscow, 123098 Russia}
\newcommand{\kyoto}{Kyoto University, Kyoto 606-8502, Japan}
\newcommand{\lawllnl}{Lawrence Livermore National Laboratory, Livermore, California 94550, USA}
\newcommand{\losalamos}{Los Alamos National Laboratory, Los Alamos, New Mexico 87545, USA}
\newcommand{\lund}{Department of Physics, Lund University, Box 118, SE-221 00 Lund, Sweden}
\newcommand{\lyon}{IPNL, CNRS/IN2P3, Univ Lyon, Universit{\'e} Lyon 1, F-69622, Villeurbanne, France}
\newcommand{\maryland}{University of Maryland, College Park, Maryland 20742, USA}
\newcommand{\mass}{Department of Physics, University of Massachusetts, Amherst, Massachusetts 01003-9337, USA}
\newcommand{\mate}{MATE, Laboratory of Femtoscopy, K\'aroly R\'obert Campus, H-3200 Gy\"ongy\"os, M\'atrai\'ut 36, Hungary}
\newcommand{\michigan}{Department of Physics, University of Michigan, Ann Arbor, Michigan 48109-1040, USA}
\newcommand{\miss}{Mississippi State University, Mississippi State, Mississippi 39762, USA}
\newcommand{\muhlenberg}{Muhlenberg College, Allentown, Pennsylvania 18104-5586, USA}
\newcommand{\nara}{Nara Women's University, Kita-uoya Nishi-machi Nara 630-8506, Japan}
\newcommand{\natmephi}{National Research Nuclear University, MEPhI, Moscow Engineering Physics Institute, Moscow, 115409, Russia}
\newcommand{\newmex}{University of New Mexico, Albuquerque, New Mexico 87131, USA}
\newcommand{\nmsu}{New Mexico State University, Las Cruces, New Mexico 88003, USA}
\newcommand{\northcg}{Physics and Astronomy Department, University of North Carolina at Greensboro, Greensboro, North Carolina 27412, USA}
\newcommand{\ohio}{Department of Physics and Astronomy, Ohio University, Athens, Ohio 45701, USA}
\newcommand{\ornl}{Oak Ridge National Laboratory, Oak Ridge, Tennessee 37831, USA}
\newcommand{\orsay}{IPN-Orsay, Univ.~Paris-Sud, CNRS/IN2P3, Universit\'e Paris-Saclay, BP1, F-91406, Orsay, France}
\newcommand{\peking}{Peking University, Beijing 100871, People's Republic of China}
\newcommand{\pnpi}{PNPI, Petersburg Nuclear Physics Institute, Gatchina, Leningrad region, 188300, Russia}
\newcommand{\pusan}{Pusan National University, Pusan 46241, Korea}
\newcommand{\riken}{RIKEN Nishina Center for Accelerator-Based Science, Wako, Saitama 351-0198, Japan}
\newcommand{\rikjrbrc}{RIKEN BNL Research Center, Brookhaven National Laboratory, Upton, New York 11973-5000, USA}
\newcommand{\rikkyo}{Physics Department, Rikkyo University, 3-34-1 Nishi-Ikebukuro, Toshima, Tokyo 171-8501, Japan}
\newcommand{\saispbstu}{Saint Petersburg State Polytechnic University, St.~Petersburg, 195251 Russia}
\newcommand{\seoulnat}{Department of Physics and Astronomy, Seoul National University, Seoul 151-742, Korea}
\newcommand{\stonybrkc}{Chemistry Department, Stony Brook University, SUNY, Stony Brook, New York 11794-3400, USA}
\newcommand{\stonycrkp}{Department of Physics and Astronomy, Stony Brook University, SUNY, Stony Brook, New York 11794-3800, USA}
\newcommand{\tenn}{University of Tennessee, Knoxville, Tennessee 37996, USA}
\newcommand{\texsu}{Texas Southern University, Houston, TX 77004, USA}
\newcommand{\titech}{Department of Physics, Tokyo Institute of Technology, Oh-okayama, Meguro, Tokyo 152-8551, Japan}
\newcommand{\tsukuba}{Tomonaga Center for the History of the Universe, University of Tsukuba, Tsukuba, Ibaraki 305, Japan}
\newcommand{\vandy}{Vanderbilt University, Nashville, Tennessee 37235, USA}
\newcommand{\weizmann}{Weizmann Institute, Rehovot 76100, Israel}
\newcommand{\wigner}{Institute for Particle and Nuclear Physics, Wigner Research Centre for Physics, Hungarian Academy of Sciences (Wigner RCP, RMKI) H-1525 Budapest 114, POBox 49, Budapest, Hungary}
\newcommand{\yonsei}{Yonsei University, IPAP, Seoul 120-749, Korea}
\newcommand{\zagreb}{Department of Physics, Faculty of Science, University of Zagreb, Bijeni\v{c}ka c.~32 HR-10002 Zagreb, Croatia}
\newcommand{\zambia}{Department of Physics, School of Natural Sciences, University of Zambia, Great East Road Campus, Box 32379, Lusaka, Zambia}
\affiliation{\abilene}
\affiliation{\augie}
\affiliation{\banaras}
\affiliation{\barc}
\affiliation{\baruch}
\affiliation{\bnlcoll}
\affiliation{\bnlphys}
\affiliation{\caucr}
\affiliation{\charlesczech}
\affiliation{\ciae}
\affiliation{\cns}
\affiliation{\colorado}
\affiliation{\columbia}
\affiliation{\czechtech}
\affiliation{\debrecen}
\affiliation{\elte}
\affiliation{\ewha}
\affiliation{\famu}
\affiliation{\fsu}
\affiliation{\gsu}
\affiliation{\hiroshima}
\affiliation{\howard}
\affiliation{\ihepprot}
\affiliation{\illuiuc}
\affiliation{\inrras}
\affiliation{\instpasczech}
\affiliation{\isu}
\affiliation{\jaea}
\affiliation{\jeonbuk}
\affiliation{\jyvaskyla}
\affiliation{\kek}
\affiliation{\korea}
\affiliation{\kurchatov}
\affiliation{\kyoto}
\affiliation{\lawllnl}
\affiliation{\losalamos}
\affiliation{\lund}
\affiliation{\lyon}
\affiliation{\maryland}
\affiliation{\mass}
\affiliation{\mate}
\affiliation{\michigan}
\affiliation{\miss}
\affiliation{\muhlenberg}
\affiliation{\nara}
\affiliation{\natmephi}
\affiliation{\newmex}
\affiliation{\nmsu}
\affiliation{\northcg}
\affiliation{\ohio}
\affiliation{\ornl}
\affiliation{\orsay}
\affiliation{\peking}
\affiliation{\pnpi}
\affiliation{\pusan}
\affiliation{\riken}
\affiliation{\rikjrbrc}
\affiliation{\rikkyo}
\affiliation{\saispbstu}
\affiliation{\seoulnat}
\affiliation{\stonybrkc}
\affiliation{\stonycrkp}
\affiliation{\tenn}
\affiliation{\texsu}
\affiliation{\titech}
\affiliation{\tsukuba}
\affiliation{\vandy}
\affiliation{\weizmann}
\affiliation{\wigner}
\affiliation{\yonsei}
\affiliation{\zagreb}
\affiliation{\zambia}
\author{N.J.~Abdulameer} \affiliation{\debrecen}
\author{U.~Acharya} \affiliation{\gsu} 
\author{A.~Adare} \affiliation{\colorado} 
\author{C.~Aidala} \affiliation{\michigan} 
\author{N.N.~Ajitanand} \altaffiliation{Deceased} \affiliation{\stonybrkc} 
\author{Y.~Akiba} \email[PHENIX Spokesperson: ]{akiba@rcf.rhic.bnl.gov} \affiliation{\riken} \affiliation{\rikjrbrc} 
\author{M.~Alfred} \affiliation{\howard} 
\author{N.~Apadula} \affiliation{\isu} \affiliation{\stonycrkp} 
\author{H.~Asano} \affiliation{\kyoto} \affiliation{\riken} 
\author{B.~Azmoun} \affiliation{\bnlphys} 
% ...
\collaboration{PHENIX Collaboration}  \noaffiliation

\date{\today}


\begin{abstract}

The invariant yield of electrons from open-heavy-flavor decays for 
$1<p_T<8$ GeV/$c$ at midrapidity $|y|<0.35$ in Au$+$Au collisions at 
$\sqrt{s_{_{NN}}}$ = 200 GeV has been measured by the PHENIX experiment 
at the Relativistic Heavy Ion Collider. A displaced-vertex analysis with 
the PHENIX silicon-vertex detector enables extraction of the fraction of 
charm and bottom hadron decays and unfolding of the invariant yield of 
parent charm and bottom hadrons. The nuclear-modification factors 
$R_{AA}$ for electrons from charm and bottom hadron decays and 
heavy-flavor hadrons show both a centrality and a quark-mass dependence, 
indicating suppression in the quark-gluon plasma produced in these 
collisions that is medium sized and quark-mass dependent.

\end{abstract}

\maketitle

\section{Introduction}
\label{sec:intro}

Charm ($c$) and bottom ($b$) quarks, with masses of 
$m_c\,{\approx}\,1.3$~\gevcsq and $m_b\,{\approx}\,4.2$~\gevcsq, are much 
heavier than the temperature reached in the quark-gluon plasma (QGP) 
produced at the Relativistic Heavy Ion Collider (RHIC) and the Large 
Hadron Collider (LHC).  As such, charm and bottom quarks, collectively 
known as heavy-flavor quarks, are produced predominantly at the 
primordial stages of high-energy nucleus-nucleus collisions and 
negligibly via interactions between thermalized particles in the QGP. 
Once produced, heavy quarks lose energy while propagating through the 
QGP and, for that reason, open-heavy-flavor hadrons are excellent probes 
of the properties of the QGP.  The current status of both experimental and 
theoretical developments is reviewed in Ref.~\cite{HFreview_2019}.

Experiments at RHIC and the LHC have measured the cross section of 
inclusive heavy flavor, as well as those for charm and bottom separated 
final states~\cite{Acharya:2018hre,Sirunyan:2018ktu,Sirunyan:2017xss,Sirunyan:2017oug,Sirunyan:2018zys,Adam_2019,Atlas_hfmu_2021,STAR_hfe_2021,ALICE_HFe_2020,ATLAS_HFmu_2018}. 
Previous measurements of separated charm and bottom heavy-flavor cross 
sections at RHIC, obtained in minimum-bias (MB) \auau collisions at \sqsn = 
200 GeV by the PHENIX Collaboration, suggest lower suppression of electrons 
from bottom hadron decays \btoe compared to those from 
charm-hadron decays (\ctoe) in the range of 
3~$<\pt<$~4~\gevc~\cite{Adare:2015hla}. This is in agreement with the widely 
postulated mass ordering for energy loss by quarks (q) and gluons (g) in the 
QGP, $\Delta E_{g} > \Delta E_{u, d, s} > \Delta E_{c} > \Delta E_{b}$ at 
$\pt>$~4~\gevc. Due to the large systematic uncertainties on the \pp 
baseline measurement, the nuclear-modification factor \raa did not 
definitively constrain the suppression pattern and mass dependence of the 
energy-loss mechanism.

Although heavy-flavor hadron-production mechanisms have been studied 
widely, the mechanisms that contribute to the in-medium modification 
thereof are not well understood. Many classes of models exist that 
employ one or more of the following effects: radiative energy 
loss~\cite{Mustafa:1997pm,Dokshitzer:2001zm}, collisional energy 
loss~\cite{Meistrenko:2012ju}, or dissociation and 
coalescence~\cite{Adil:2006ra} of heavy-flavor hadrons in the medium. 
While radiative energy loss is significant at high 
\pt~($>\,\approx\,$10~\gevc), theoretical models suggest that collisional 
energy loss is equally important at low \pt~\cite{Adil:2006ra}. 
Cold-nuclear-matter effects, such as the Cronin effect for heavy 
quarks, could also play an important role in the interpretation of these 
observations at low to medium \pt~\cite{Adare:2012yxa}. For these 
reasons, a precise measurement of the nuclear-modification factor \raa 
over a broad range of momentum and centrality is necessary to 
investigate the interplay between competing mechanisms that could 
contribute to the suppression or enhancement seen in different regions 
of phase space.

This paper reports on the measurement of electrons from semileptonic 
decays of open charm and bottom hadrons at midrapidity \midy in \auau 
collisions at \sqsn = 200 GeV. Using the combination of the 
high-statistics data set recorded in 2014 and the updated \pp reference 
from 2015~\cite{Aidala_2019}, nuclear-modification factors \raa of 
separated charm and bottom electrons in MB \auau as well as four 
centrality classes in \auau can be measured with improved precision 
compared to our previously published results~\cite{Adare:2015hla}.

This paper is organized as follows: Section~\ref{sec:setup} provides a 
brief introduction to the PHENIX detector, with special emphasis on the 
central arm detectors pertinent to this measurement. 
Section~\ref{sec:ana} details track reconstruction, electron 
identification, event selection, background estimation, signal 
extraction, and unfolding. Section~\ref{sec:sys} describes 
systematic-uncertainty estimates. Section~\ref{sec:result} provides the 
results of the measurement, along with comparisons with theoretical 
models. Finally, Section~\ref{sec:summary} gives the summary and 
conclusions.


\section{Experimental Setup}
\label{sec:setup}

PHENIX has previously published the decay-electron contribution from 
charm and bottom decays separately~\cite{Adare:2015hla, Aidala_2019} 
through the combination of electron-identification detectors in the 
central arms covering \midy, and the measurement of event-vertex and 
decay-electron trajectories provided by an inner silicon tracker (VTX). 
The detector systems relevant to this measurement are discussed below, 
while a detailed description of the PHENIX detector is given in 
Refs.~\cite{Adcox:2003zm,osti_15007383,mannel_vtx}.

The VTX is described in detail in Refs.~\cite{ Adare:2009ic, 
Aidala_2019}. It is composed of two arms, each with $|\eta|<1$ and 
$\Delta\phi\,{\approx}\,0.8\pi$ coverage. Each arm has four layers around the 
beam pipe. The radial distances of these layers from the nominal beam 
center are 2.6, 5.1, 11.8, and 16.7~cm. The innermost two 
layers have pixel segmentation of 50~$\times$~425~$\mu$m. The two 
outer layers have strip segmentation of 80~$\times$~1000~$\mu$m.

\section{Analysis Method}
\label{sec:ana}

This paper reports measurements using data collected by the PHENIX 
experiment during the 2014 high-luminosity \auau collisions at \sqsn = 
200 GeV. The data were recorded with a MB trigger and 
correspond to an integrated luminosity of 2.3 ${\rm nb^{-1}}$. A set of 
event, offline track and electron selection cuts were applied as 
described below.

\subsection{Event selection}

Events considered here are characterized by the MB trigger, which 
requires simultaneous activity in both beam-beam-counter (BBC) phototube 
arrays located at pseudorapidity $3.0<|\eta|<3.9$ and the 
zero-degree-calorimeter at 18~m downstream from the intersection point. 
This criterion selects $93{\pm}2\%$ of the \auau inelastic cross 
section. The total number of charged particles as measured by the BBC 
determines the collision centrality.  The BBC is also used later to 
calculate the number of nucleon participants and the number of binary 
collisions via comparisons with Monte-Carlo-Glauber model simulations of 
the collisions~\cite{ppg014}. The results shown here are for MB Au$+$Au 
collisions and 0\%--10\%, 10\%--20\%, 20\%--40\% and 40\%--60\% 
centrality classes.

The collision vertex is determined by clusters of converging VTX tracks. 
The vertex resolution is determined from the standard deviation of the 
difference between the vertex position measured by each VTX at the east 
and west arm. The vertex resolutions for $x-y-z$ coordinate are 
$(\sigma_x, \sigma_y, \sigma_z) = (44, 38, 48)~\mu{\rm m}$. The radial 
beam profile during the 2014 run had a width of 45~$\mu$m and was very 
stable during beam fills. The beam-center position in the $xy$ plane was 
then determined from the average position during the fill to avoid 
autocorrelations between the vertex determination and the distance of 
closest approach (\dca) measurements in each event. Because of the 
modest RHIC collision rates in 2014 of less than 10 kHz in \auau 
collisions, no significant contributions were found of multiple 
collisions per beam crossing or signal pileup in the dataset. The 
analysis required a z-vertex within $\pm$10~cm reconstructed by the 
VTX detector.

\subsection{Track Reconstruction}

Charged-particle tracks are reconstructed (trajectory and momentum) by 
the PHENIX central-arm drift chambers (DC) and pad chambers covering the 
pseudorapidity $|\eta|<0.35$ and azimuthal angle $\Delta\phi=\pi/2$. To 
identify electrons and positrons, the reconstructed tracks are projected 
to the ring-imaging \cherenkov Detector (RICH).  Electrons and positrons 
are collectively referred to here as electrons.  In the momentum 
range where charged pions are below the RICH radiator threshold 
($p_T<4.7$~GeV/$c$), tracks are required to be associated with signals in 
two phototubes within a radius expected of electron \cherenkov rings.  
Above this threshold, to aid in eliminating pion background, associated 
signals in three phototubes are required.  Additional tracking 
information is provided by pad chambers that are immediately behind the 
RICH.

Energy-momentum matching is also required for electron identification. 
Electromagnetic calorimeters (EMCal) are the outermost detectors in the 
PHENIX central arms. The EMCal comprises eight sectors, two of 
which are lead-glass layers, and six of which are lead-scintillator 
layers.  Tracks with measured momentum $p$ that are associated with 
showers in the calorimeters of energy $E$ are characterized by the 
variable \dep=($E/p-\mu_{E/p}$)/$\sigma_{E/p}$, where $\mu_{E/p}$ and 
$\sigma_{E/p}$ are the mean and standard deviation of a precalibrated 
Gaussian $E/p$ distribution. The requirement of \dep$>-2$ further 
removes background from hadron tracks associated with \cherenkov rings 
produced by nearby electrons or high-momentum pions. Remaining 
background contributions are quantified as discussed below.


\begin{figure}[htb] 
\includegraphics[width=1.0\linewidth]{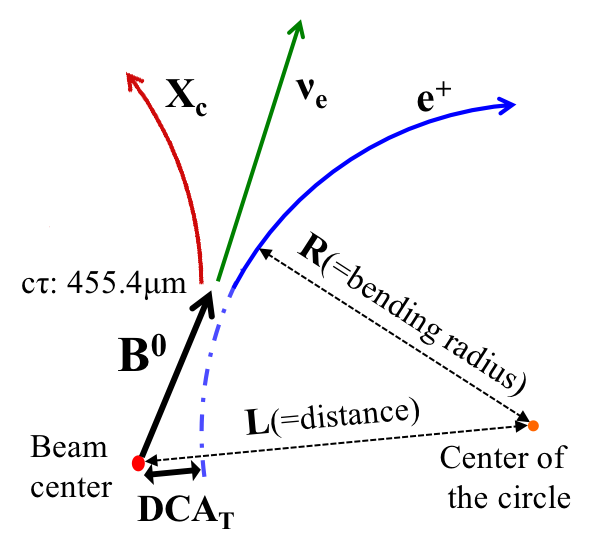}
\caption{Definition of the distance of the closest approach \dcat in the 
transverse plane (normal to the beam direction).}
\label{fig:dcat}
\end{figure}

The reconstructed tracks are then associated to VTX hits to perform the 
displaced tracking around the collision vertex. Taking advantage of the 
different decay lengths of charm and bottom hadrons (viz.~for $D^0$ the 
decay length is $c\tau$=~122.9~$\mu$m and for the $B^0$ it is 
$c\tau=$~455.4~$\mu$m~\cite{ProgTheoExpPhys98030001}), electrons from 
these decays are statistically separated based on the \dcat in the 
transverse plane ($x$--$y$, normal to the beam direction) to the 
collision vertex.  Figure~\ref{fig:dcat} illustrates the definition of 
\dcat~=~$\mathbf{L-R}$ for a VTX-associated track, where $\textbf{R}$ is 
a radius of the circle defined by the track trajectory in the constant 
magnetic field around the VTX region and $\textbf{L}$ is the length 
between the beam center and the center of the circle.

\subsection{Background estimation}
\label{subsec:bg}

\subsubsection{Misreconstruction}
\label{subsec:misreco}

In a high-multiplicity environment, tracks are accidentally 
reconstructed with hits from different particles. Misreconstructed 
tracks have two sources: (i) misidentified hadrons composed of tracks 
accidentally matching RICH \cherenkov rings or EMCal clusters; and (ii) 
mismatches between DC tracks and uncorrelated VTX hits.

\begin{figure}[htb]
\includegraphics[width=1.0\linewidth]{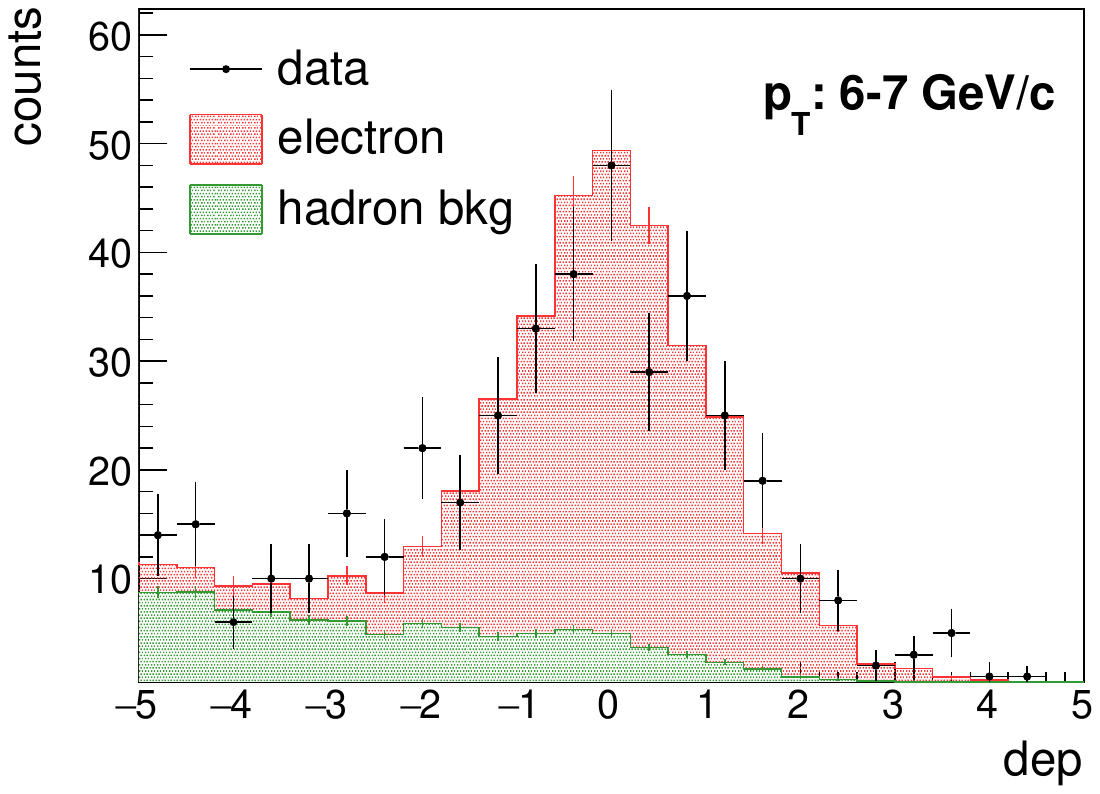}
\caption{A fit result of the \dep distribution for electron candidates 
with \pt = 6--7~\gevc in MB Au$+$Au collisions. The red and 
green distributions are the estimated contributions for electrons and 
hadron backgrounds.}
\label{fig:dep_tmpfit}
\end{figure}

The misidentified hadron-track contamination is estimated with a sample 
of tracks where the sign of their $z$-direction is swapped. The swapped 
tracks that, after being projected to RICH, match \cherenkov rings 
provides the expected number of misidentified hadrons.  Charged hadrons 
with momentum $p>4.7$~\gevc also radiate \cherenkov light and make 
RICH hits, meaning the swap method underestimates the fraction of 
misidentified hadrons.  
The contamination at high \pt is estimated by the \dep template method, 
in which the measured \dep distribution is assumed to be the sum of the 
electron distribution and the hadron-background distribution. The \dep 
template for the electron distribution is obtained by the RICH swap 
method for $\pt \leq 4.5$~\gevc, where the hadron contamination is very 
small. The \dep template for hadron backgrounds is obtained by vetoing 
the electron candidates from all reconstructed tracks. The measured \dep 
distribution for $\pt>4.5$~\gevc is fitted with the electron and hadron 
background templates. An example of the \dep template method is shown in 
Fig.~\ref{fig:dep_tmpfit} for electron candidates at $6<\pt<7$~\gevc 
in MB Au$+$Au collisions. The electron signal in the \dep distribution 
is centered at \dep = 0. The background tail due to hadrons overlaps the 
signal region. The hadron background increases at higher \pt.

The mismatch between DC tracks and uncorrelated VTX hits is estimated by 
the VTX swap method, which intentionally creates a mismatch by changing 
the angle of DC tracks by 10 degrees in the $\phi$--$\eta$ plane. The 
10-degree rotation is sufficiently larger than the angular resolution of 
the DC such that the rotated tracks are never connected with VTX hits 
belonging to the same particle.

\subsubsection{Photonic background}

Photonic electrons are the main background source in this analysis. They 
are produced by internal conversions (Dalitz decay) and photon 
conversions at the beam pipe and the first VTX layer. Photonic 
conversions produced in the other layers of the VTX do not produce 
tracks accepted by the tracking algorithm because the presence of a hit 
in the first layer is required. Electron pairs from converted photons 
have a small opening angle, therefore it is required that an electron 
track should not have a neighboring electron track with 
$-0.02<$~\textit{\rm chrg}~$\times\Delta\phi<0.04$ radian for 
$\pt<1.8$~\gevc and narrower for high \pt, where \textit{\rm chrg} is the 
charge of the track and $\Delta\phi$ is the azimuthal difference of 
electron pairs.  This isolation cut minimizes the contamination from 
internal and external conversion electrons, and is the same as described 
in Ref.~\cite{Adare:2015hla}.

The number of electrons obtained after removing background from 
misidentified and mismatched tracks but before the isolation cut, 
($N_e$), is the sum of photonic ($N_P$) and nonphotonic sources 
($N_{\rm NP}$):
\begin{eqnarray}
 N_e = N_P + N_{\rm NP}, \label{eq:ne}
\end{eqnarray}
while the number of electrons after the isolation cut is
\begin{eqnarray}
 \tilde{N}_e = \varepsilon_P \times \varepsilon_{\rm UC} 
\times N_P + \varepsilon_{\rm UC} \times N_{\rm NP}, \label{eq:ne_v}
\end{eqnarray}
where $\varepsilon_P$ is the survival rate after the isolation cut for 
the correlated pairs such as photonic electrons, and 
$\varepsilon_{\rm UC}$ is the survival rate for the uncorrelated tracks. 
The $\varepsilon_{\rm UC}$ is also applied to both the photonic and 
nonphotonic electrons because uncorrelated tracks appear everywhere. By 
solving Eqs.~\eqref{eq:ne} and \eqref{eq:ne_v} simultaneously, $N_P$ and 
$N_{\rm NP}$ are described as
\begin{eqnarray}
 N_P = \frac{\tilde{N}_e - N_e \varepsilon_{\rm UC}}{\varepsilon_{\rm UC} ( \varepsilon_P -1)},
\end{eqnarray}
and
\begin{eqnarray}
 N_{\rm NP} = \frac{N_e  \varepsilon_P \varepsilon_{\rm UC} - \tilde{N}_e }{\varepsilon_{\rm UC} ( \varepsilon_P -1)}.
\end{eqnarray}
The fraction of photonic and nonphotonic electrons is then written as
\begin{eqnarray}
 F_{P} = \frac{\varepsilon_P \varepsilon_{\rm UC} N_{P}}
{\varepsilon_P \varepsilon_{\rm UC} N_{P} + \varepsilon_{\rm UC} N_{\rm NP}},
\end{eqnarray}
and
\begin{eqnarray}
 F_{\rm NP} = \frac{\varepsilon_{\rm UC} N_{\rm NP}}
 {\varepsilon_P \varepsilon_{\rm UC} N_{P} + \varepsilon_{\rm UC} N_{\rm NP}}.
\end{eqnarray}

\begin{figure}[htb]
\includegraphics[width=1.0\linewidth]{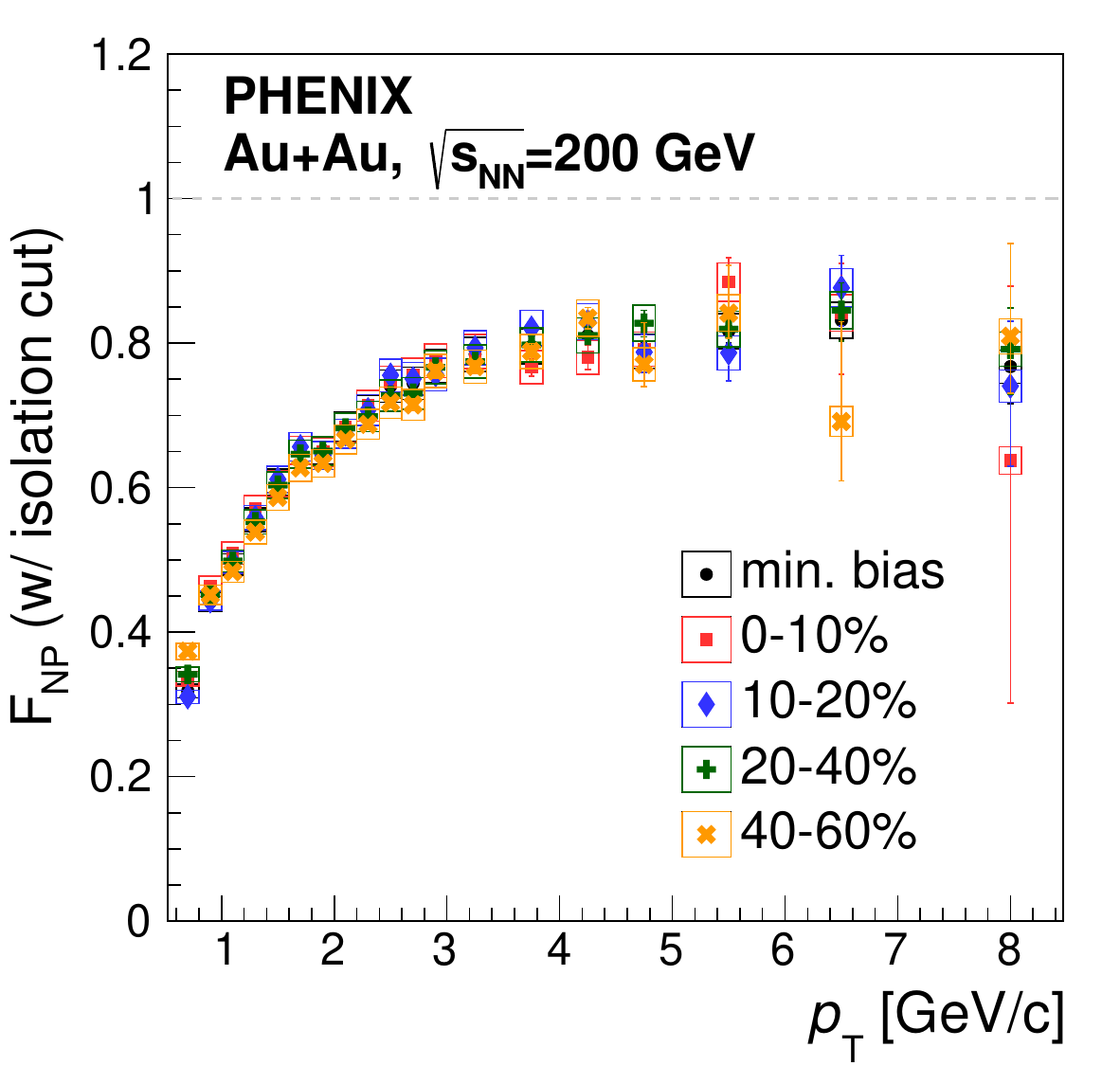}
\caption{The fraction of nonphotonic electrons ($F_{\rm NP}$) as a 
function of \pt for MB and the indicated four centrality classes. }
\label{fig:FNP}
\end{figure}

Figure~\ref{fig:FNP} shows $F_{\rm NP}$ as a function of \pt for MB 
\auau collisions as well as four centrality classes, which correspond to 
0\%--10\%, 10\%--20\%, 20\%--40\% and 40\%--60\%.  The $F_{\rm NP}$ 
values increase with $p_T$ and their curves are similar for all 
centrality classes.

\subsubsection{Nonphotonic background}

\begin{figure}[htb]
\includegraphics[width=1.0\linewidth]{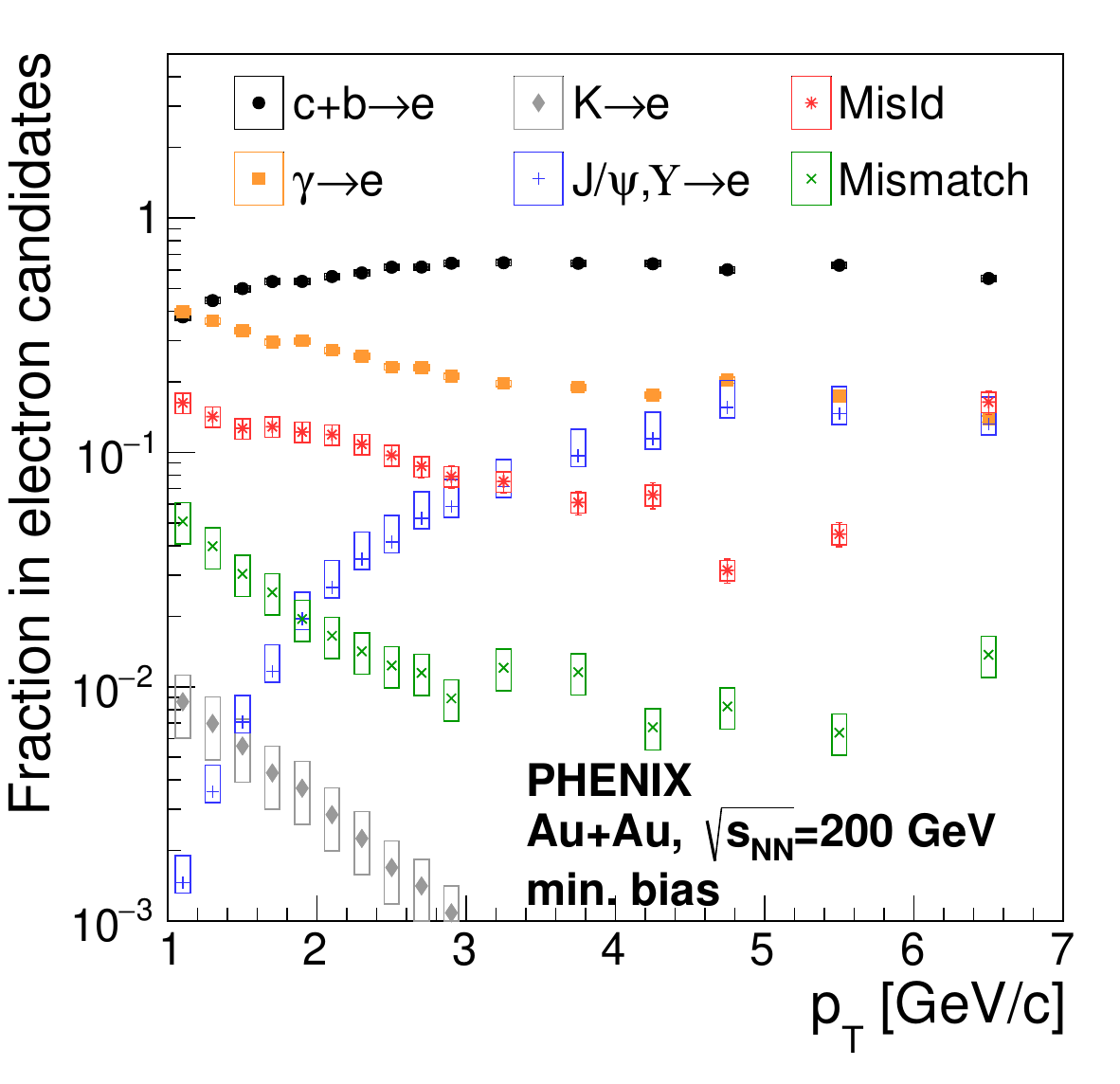}
\caption{The fractions of signal component in isolated electron-track 
candidates as a function of \pt in MB \auau collisions. The 
isolation cut is applied. The modeling of these backgrounds is described 
in the text and in Ref.~\cite{Adare:2015hla}.}
\label{fig:Frac_e}
\end{figure}

Nonphotonic background sources are electrons from the three-body decays 
of kaons and the decay of J/$\psi$ and $\Upsilon$. The other 
contributions from the resonance decays of $\rho, \omega$, $\phi$ and 
the Drell-Yan process are found to be negligibly small compared to the 
total background. The nonphotonic backgrounds included in $F_{\rm NP}$ 
are estimated by the full \geant simulation of the PHENIX detector with 
measured particle yields~\cite{PHENIX_PIDref, PHENIX_jpsref} as inputs 
and normalized by the background cocktail, applying with the 
uncorrelated survival rate $\varepsilon_{\rm UC}$. The detailed modeling of 
these backgrounds is described in Ref.~\cite{Adare:2015hla}.  After 
subtracting these backgrounds, the remaining signal component is the 
inclusive heavy flavor ($F_{c+b}$).
Figure~\ref{fig:Frac_e} shows the fractions of signal, photonic, and 
nonphotonic backgrounds of isolated electrons in MB \auau 
collisions.

\subsection{Invariant yields of heavy-flavor electrons}
\label{sec:hfeyield}

\begin{figure}[htb]
\includegraphics[width=1.0\linewidth]{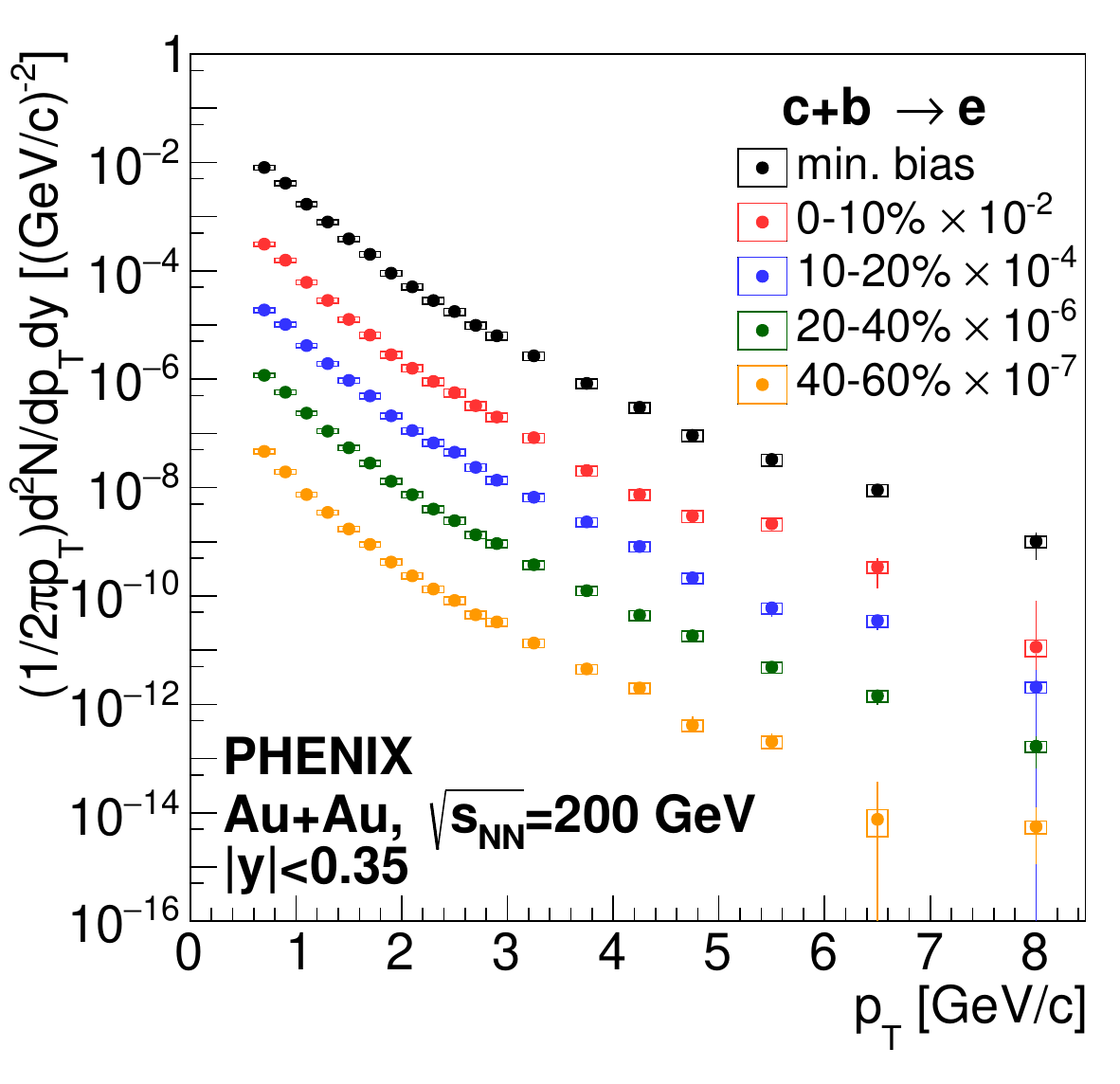}
\caption{The invariant yields of \cbtoe as a function of \pt for 
different \auau centrality classes. These spectra are scaled by factors 
of 10 for clarity.}
\label{fig:InvYield_cbe}
\end{figure}

The invariant yield of heavy-flavor electrons is calculated from the 
photonic electron yields and the fraction of heavy-flavor electrons to 
photonic electrons as
\begin{eqnarray}
\frac{d^{2}N^{c+b}_{e}}{dp_{T}dy} = \frac{d^2 N_e^{c+b} (N_e^{\gamma}) }{dp_t dy} \times \frac{F_{c+b}}{F_P},
\end{eqnarray}
where $N_e^{c+b}$~($N_e^{\gamma}$), $F_{c+b}$~($F_P$), and 
$d^2N_e^{\gamma}/dp_Tdy$ are the yield, fraction, and invariant yield, 
respectively, of heavy-flavor (photonic) electrons.  The photonic 
electron yield is calculated based on the invariant yields of $\pi^0$ 
and $\eta$ measured by PHENIX~\cite{cocktail_pi0ref,cocktail_etaref}, 
using a method which has been demonstrated to give an accurate 
description of photonic electron yields in the previous heavy-flavor 
electron measurement~\cite{Adare:2010de,Adare:2015hla}. The fractions 
$F_{c+b}$ and $F_P$ are determined by the data-driven method described 
in the previous section.  Note that the efficiency and acceptance cancel 
out in $F_{c+b}$ and $F_P$. The invariant yields of heavy-flavor 
electrons (\cbtoe) in MB \auau as well as four centrality classes in 
\auau are shown in Fig.~\ref{fig:InvYield_cbe}. The bars and boxes 
represent statistical and systematic uncertainties which are described 
in Section~\ref{sec:sys}.

\subsection{\dcat distribution of the background}
\label{sec:dcabg}

\begin{figure}[htb]
\includegraphics[width=1.0\linewidth]{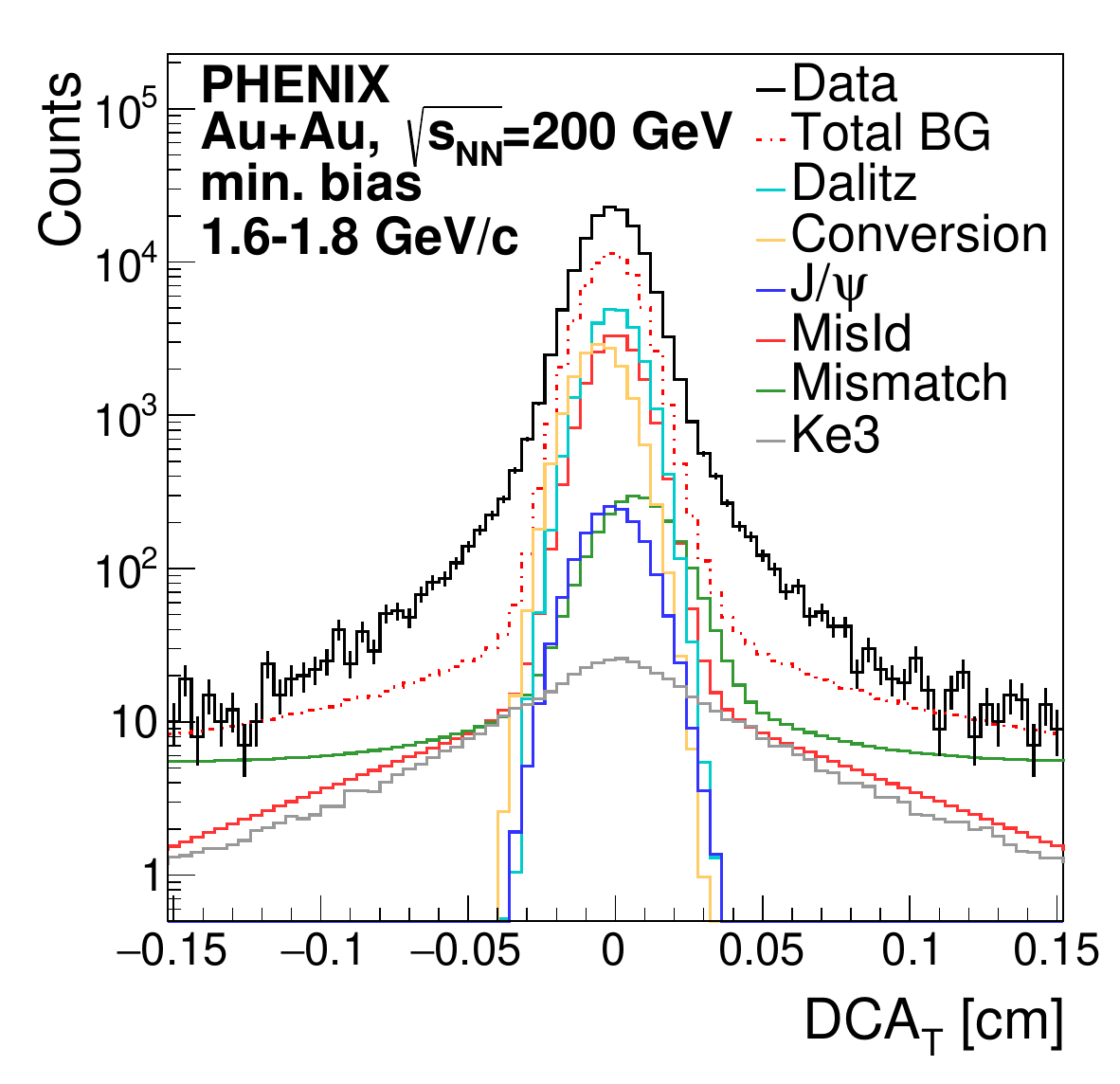}
\caption{\dcat distribution of electron candidates for 1.6 $<p_T<$ 
1.8~\gevc in MB collisions. All background components are also 
plotted.}
\label{fig:DcaBg}
\end{figure}

The \dcat distribution of misidentified hadrons and mismatched backgrounds 
are determined by the RICH and VTX swap method as described in 
Section~\ref{subsec:misreco}. The swap method is data driven and the 
obtained \dcat distribution includes the normalization and resolution 
effects. Photonic- and nonphotonic-background \dcat distributions are 
determined by the full \geant simulation of the PHENIX detector. 
Background sources are generated with the \pt distribution measured by 
PHENIX and decay electron tracks are reconstructed and analyzed with the 
same analysis cuts used to calculate \dcat. The obtained \dcat 
distributions are fitted with Gaussian functions for photonic, \jps, and 
$\Upsilon$ backgrounds, and Laplace functions for kaon backgrounds to 
obtain smooth shapes. These \dcat distributions are normalized by the 
factors described in the previous section (\ref{subsec:misreco}).

The \dcat resolution of the data and the Monte-Carlo simulation are 
compared. The resolution of the \dcat distribution is a convolution of 
the position resolution of the VTX and the beam spot size.  The 
simulation was generated with ideal VTX geometry and a single 
beam-spot-size value and smeared to correct for differences with the real data 
caused by irreducible misalignments including the time dependence of the 
beam spot size during data taking.  The smearing is calculated as a 
function of \pt by comparing the \dcat width of charged hadrons between 
data and simulation.  The smearing is independent of the collision 
centrality because \dcat is measured from the beam center.

Figure~\ref{fig:DcaBg} shows the smeared and normalized \dcat 
distributions for these background sources.  Most of the background 
sources are primary particles showing up in the \dcat distributions as 
Gaussian shapes. Kaon-decay electrons as well as misidentified and 
mismatched backgrounds have large \dcat tails. Misidentified hadrons 
contain long-lived hadrons such as $\Lambda$ particles causing large 
\dcat tails. Mismatch tracks also cause large tails in the \dcat 
distribution because they are formed by hits from different particles.

\subsection{Unfolding}
\label{subsec:unfold}

\begin{figure}[htb]
\includegraphics[width=1.0\linewidth]{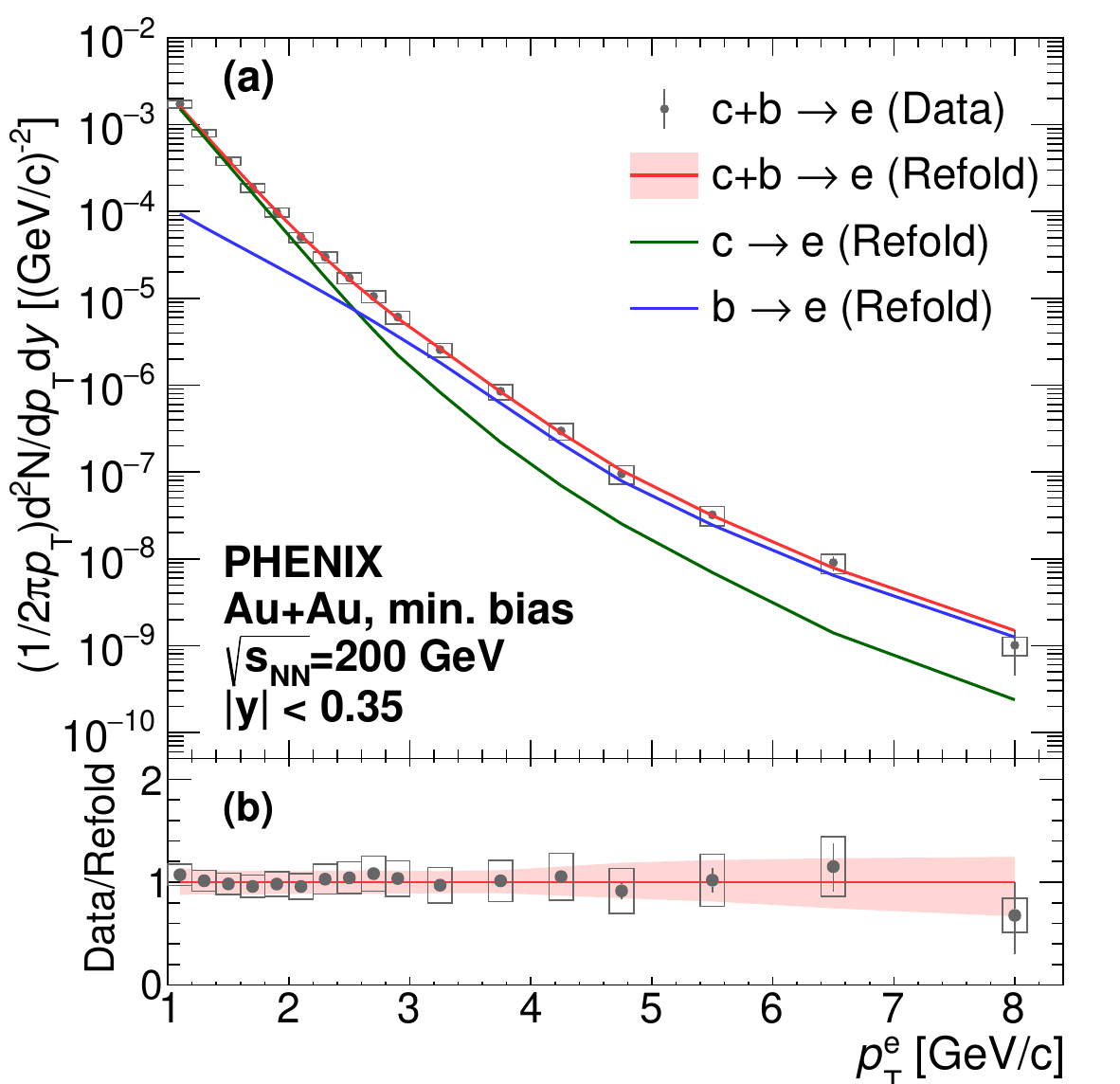}
\caption{The measured invariant yield for (black markers) \cbtoe 
as a 
function of \pt and refolded yields for (red line) \cbtoe, 
(green line) \ctoe, and (blue line) \btoe in MB \auau collisions.}
\label{fig:RefoldYield}
\end{figure}

\begin{figure}[htb]
\includegraphics[width=1.0\linewidth]{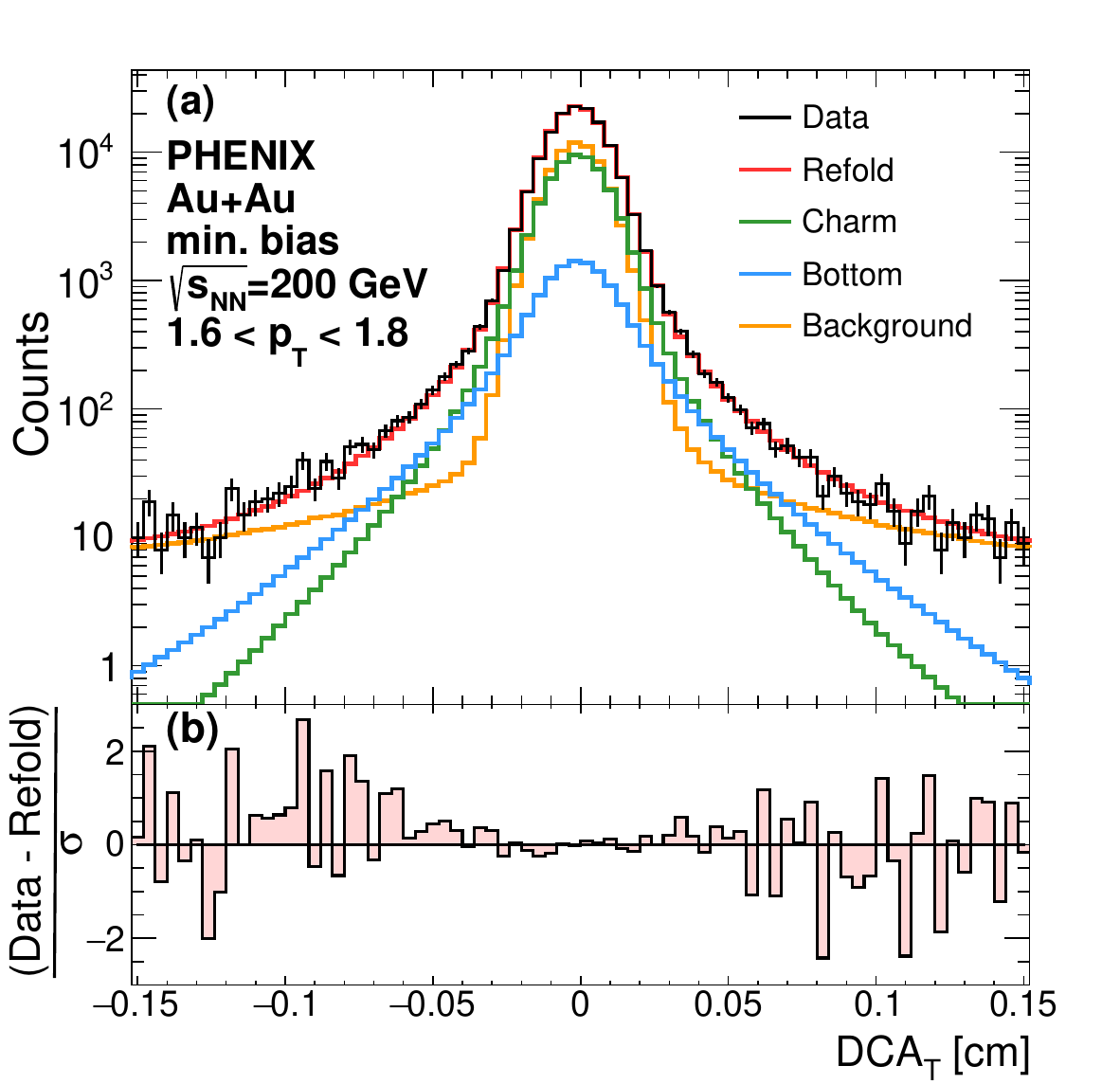}
\caption{The measured \dcat distribution of ([black] line) electron tracks, 
([red] line) refolded \cbtoe, ([yellow] line) background, ([green] 
line)\ctoe, and ([blue] line) \btoe in MB \auau collisions for 
$1.6<\pt<1.8$~\gevc.}
\label{fig:RefoldDca}
\end{figure}

\begin{figure}[htb]
\vspace{-1.3cm}
\includegraphics[width=1.0\linewidth]{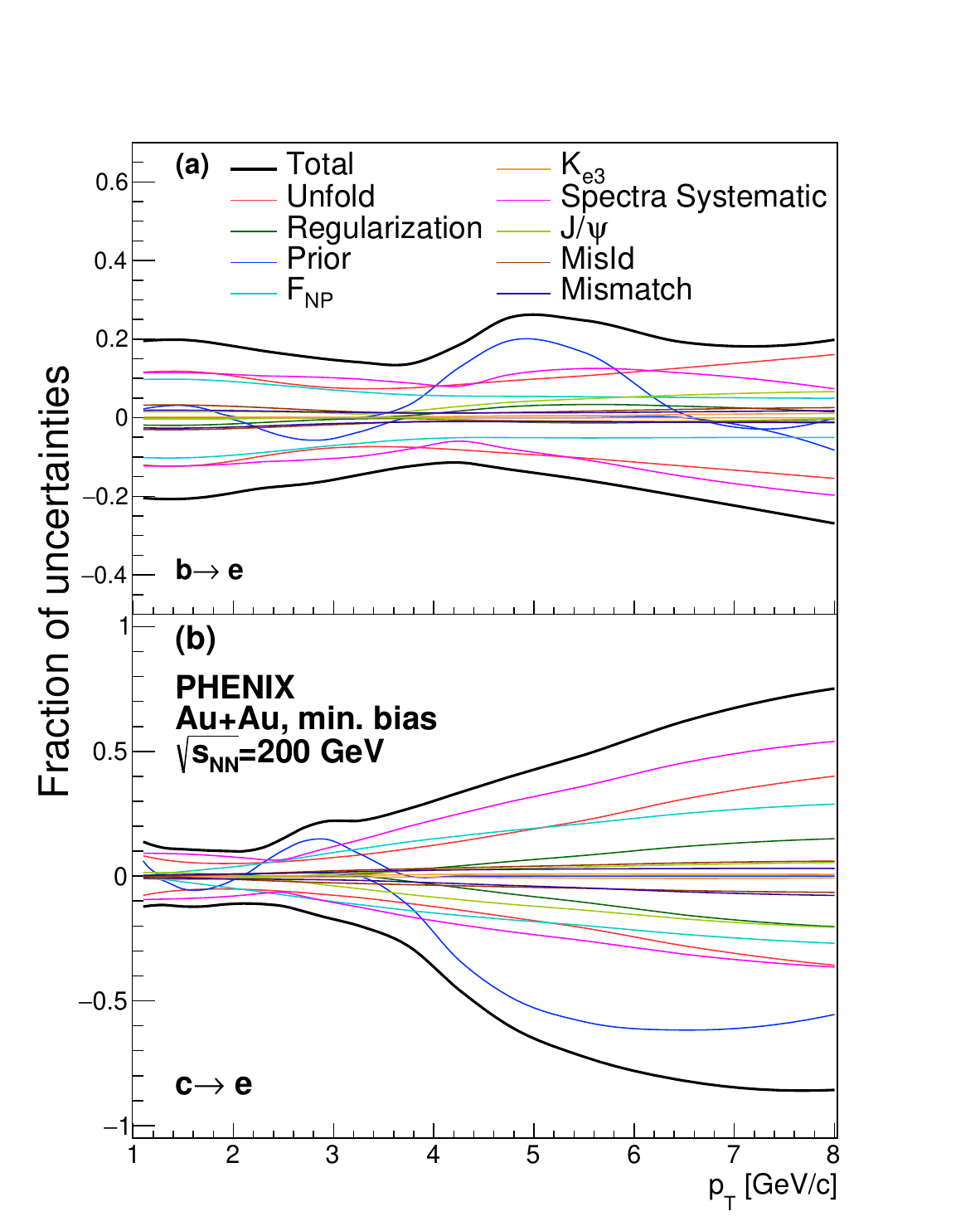}
\vspace{-0.4cm}
\caption{The fractions of each systematic uncertainty in the invariant 
yield of (a) \btoe and (b) \ctoe (in MB \auau collisions.}
\label{fig:SysErr}
\end{figure}

Because the \pt spectra and decay lengths of charm and bottom hadrons 
are significantly different, simultaneous fits to the \pt and \dcat 
distributions of heavy-flavor electrons enable separation of \ctoe and 
\btoe components. However, the \pt and \dcat template distributions for 
\ctoe and \btoe depend on unmeasured \pt spectra of the parent charm and 
bottom hadrons. To solve this inverse problem and to measure the hadron 
yields, the decay of heavy-flavor hadrons into final-state electrons is 
characterized by using a Bayesian-inference unfolding method that was 
also used by PHENIX in previous 
publications~\cite{Adare:2015hla,Aidala_2019}.

This unfolding procedure is a likelihood-based approach that uses the 
Markov-chain Monte-Carlo (MCMC) 
algorithm~\cite{Foreman_Mackey_2013_mcmc} to sample the parameter space 
and maximize the joint posterior probability distribution. The response 
matrix or decay matrix assigns a probability for a hadron at given 
$\pt^{h}$ to decay into an electron with $\pt^{e}$ and \dcat. The yields 
of charm and bottom hadrons with 17 \pt bins each within 
$0<\pt^{h}<20$~\gevc are set as unfolding parameters.

The {\sc pythia}6 generator\footnote{Using {\sc pythia}6.2 with CTEQ5L parton 
distribution function, the following parameters were modified: MSEL=5, 
MSTP(91)=1, PARP(91)=1.5, MSTP(33)=1, PARP(31)=2.5. For bottom 
(charm) hadron studies, PARJ(13)=0.75(0.63), PARJ(2)=0.29(0.2), 
PARJ(1)=0.35(0.15).}~\cite{Sjostrand:2006za_pythia} is used to model the 
decay matrix, which includes charm ($D^0, D^{\pm}, D_s, 
\Lambda_c$), and bottom hadrons ($B^0, B^{\pm}, B_s, \Lambda_b$) from the 
whole rapidity range decaying into electrons within $|y| < $ 0.35. The 
relative contributions of the charm hadrons and bottom hadrons are 
modeled by {\sc pythia}. Thus, the decay matrix has some model dependence 
which may affect the final results. 

In the decay matrix, there are two assumptions. One is that the rapidity 
distributions of hadrons are not changed in $A$$+$$A$ collisions. The BRAHMS 
collaboration reported~\cite{BRAHMS_HADRAA} that the nuclear 
modification of pions and protons at $y\,{\approx}\,3$ is similar to that at 
midrapidity. The rapidity modification is also less sensitive to the 
final result because electron contributions from large rapidity to the 
PHENIX acceptance with $|y|<$ 0.35 are small. The second assumption is 
that the relative contributions of charm (bottom) hadrons are unchanged. 
The charm hadrons have their own decay lengths which can affect the 
final results. Charm-baryon enhancement in Au$+$Au collisions was 
reported by the STAR collaboration~\cite{STAR_Lambdac_enhance}. To study 
the effect of this, the baryon enhancement for charm and bottom hadrons 
was tested using a modified decay 
matrix~\cite{nagashima_kazuya_2021_4745975}. Following 
Ref.~\cite{BaryonEnhance}, the baryon enhancement for charm and bottom 
is assumed to be the same as that for strange hadrons. The result is 
that baryon enhancement produces a lower charm-hadron yield and a higher 
bottom-hadron yield at high \pt , but the difference is within the 
systematic uncertainties discussed in the next section. The test result 
is not included in the final result.

In each sampling step, a set of hadron yields are selected by the MCMC 
algorithm. The \pt and \dcat distributions in the decay-electron space 
are predicted by applying corresponding decay matrices to the sampled 
values. The predicted \pt and \dcat distributions along with the 
measured ones are used to compute a log-likelihood:
\begin{equation}
\ln{\mathcal{L}} = \ln{P(\bm{Y}^{{\rm data}}|\bm{Y(\theta))}} +   \sum^{12}_{j=1}\ln{P(\bm{D}^{{\rm data}}_j | \bm{D_j(\theta)})}
\end{equation}
where $\bm{Y}^{{\rm data}}$ and $\bm{D}^{{\rm data}}_{{\rm j}}$ 
represent a vector of measured \pt and 12 vectors of measured \dcat in 
the range of 1.0--8.0 and 1.6--6.0~\gevc, respectively. For the 
40\%--60\% centrality bin, 11 vectors of measured \dcat in 
1.6--5.0~\gevc are used due to statistical limitations. The $\bm{Y}({\rm 
\theta})$ and $\bm{D}({\rm \theta})_{{\rm j}}$ represent the \pt and 
\dcat distribution predicted by the unfolding procedure.  MCMC repeats 
the process through multiple iterations until an optimal solution is 
found. Only statistical uncertainties in the data are included in the 
calculation of the log-likelihood.

The analyzing power to separate charm and bottom contributions is mainly 
contained in the tail of the \dcat distribution, but the \dcat 
distribution has a sharp peak with many measurements at \dcat=~0, which 
dominates the likelihood calculation in the unfolding method. A 
5\% uncertainty is added in quadrature to the statistical 
uncertainty when a given \dcat bin has a yield above a threshold that
was set to 100.

Without additional information, the unfolding procedure introduces 
large statistical fluctuations in the unfolded distributions due to 
negative correlations of adjacent bins. However, the unknown hadron 
spectra are expected to be relatively smooth. This prior belief of 
smoothness, $\pi$, is multiplied with the likelihood to get a posterior 
distribution $P$ as
\begin{equation}
    \ln{\pi(\theta)} = -\alpha^{2}(|\bm{LR}_{c}|^{2} + |\bm{LR}_{b}|^{2}),\label{eq:10}
\end{equation}
and
\begin{equation}
\ln{P} =  \ln{\mathcal{L}} + \ln{\pi(\theta)},
\end{equation}
where $\bm{L}$ denotes a 17$\times$17 matrix of regularization 
conditions and, $\bm{R}_b(\bm{R}_c)$ is the ratio of the trial bottom 
(charm) spectra to the prior. The strength of regularization is 
characterized using a parameter $\alpha$ that is tuned by repeating the 
unfolding procedure with several values of $\alpha$ and selecting the 
one that gives a maximum of the posterior distribution.

Once the unfolded charm- and bottom-hadron \pt spectra are obtained, the 
same response matrices are applied to the heavy-flavor hadron 
distribution to obtain refolded \cbtoe yields. 
Figure~\ref{fig:RefoldYield} shows the refolded invariant yield of 
\cbtoe compared to the measured data, which is in reasonable agreement 
with the refolded spectrum. Figure~\ref{fig:RefoldDca} compares the 
refolded \dcat distributions to the measured data.  The \dcat 
distribution is fit with the refolded components within 
$|\dcat|<0.1$~cm, and indicates good agreement between the measured and 
refolded distributions.

\section{Systematic uncertainties}
\label{sec:sys}

\begin{figure*}[htb]
\begin{minipage}{0.48\linewidth}
\includegraphics[width=0.99\linewidth]{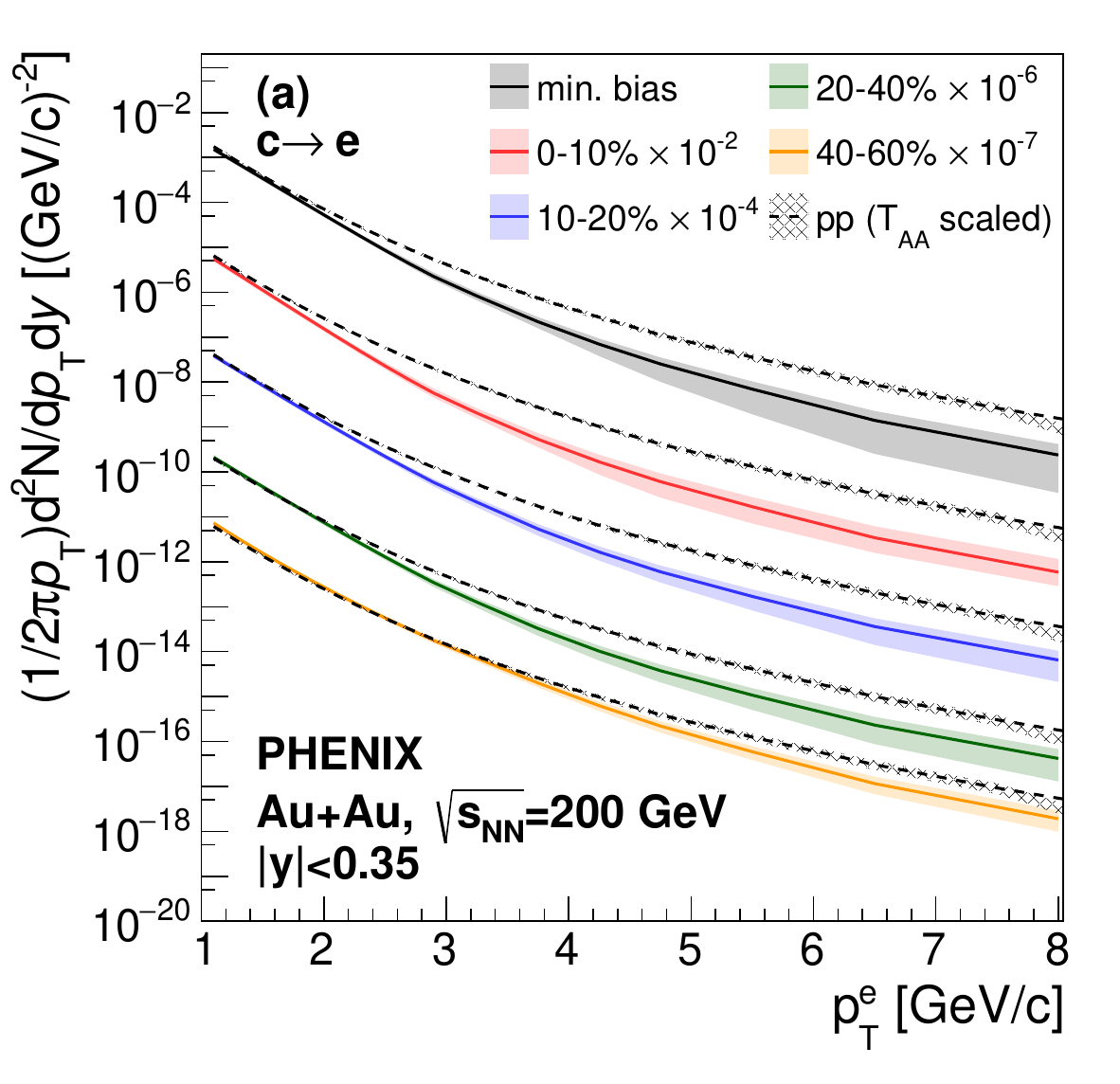}
\includegraphics[width=0.99\linewidth]{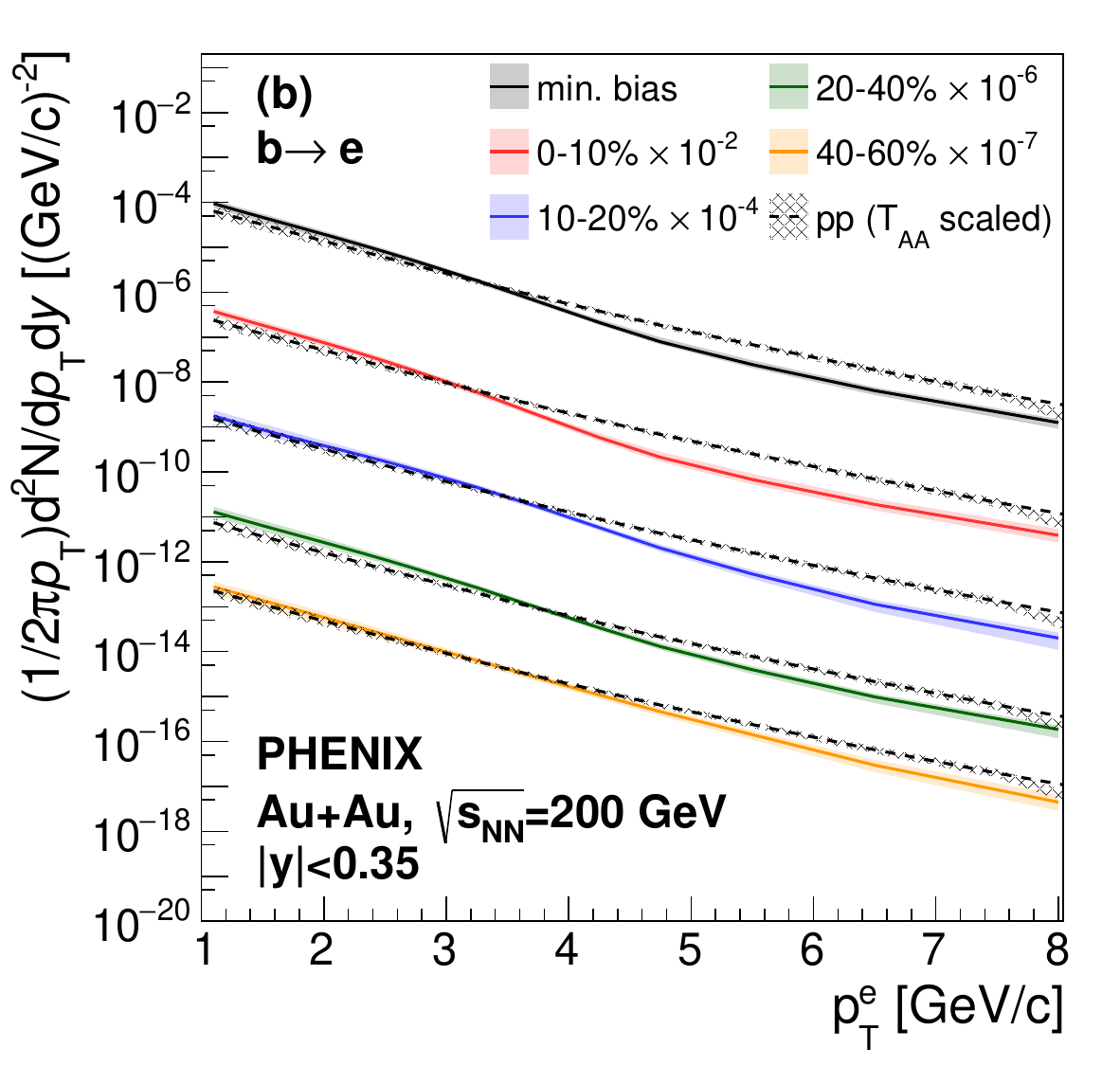}
\caption{Comparison of refolded \pt spectrum of (a) \ctoe and 
(b) \btoe in \auau collisions to that scaled by $T_{AA}$ in \pp 
collisions~\cite{Aidala_2019}.
}
\label{fig:InvYield_cebe}
\end{minipage}
\hspace{0.4cm}
\begin{minipage}{0.48\linewidth}
\includegraphics[width=0.99\linewidth]{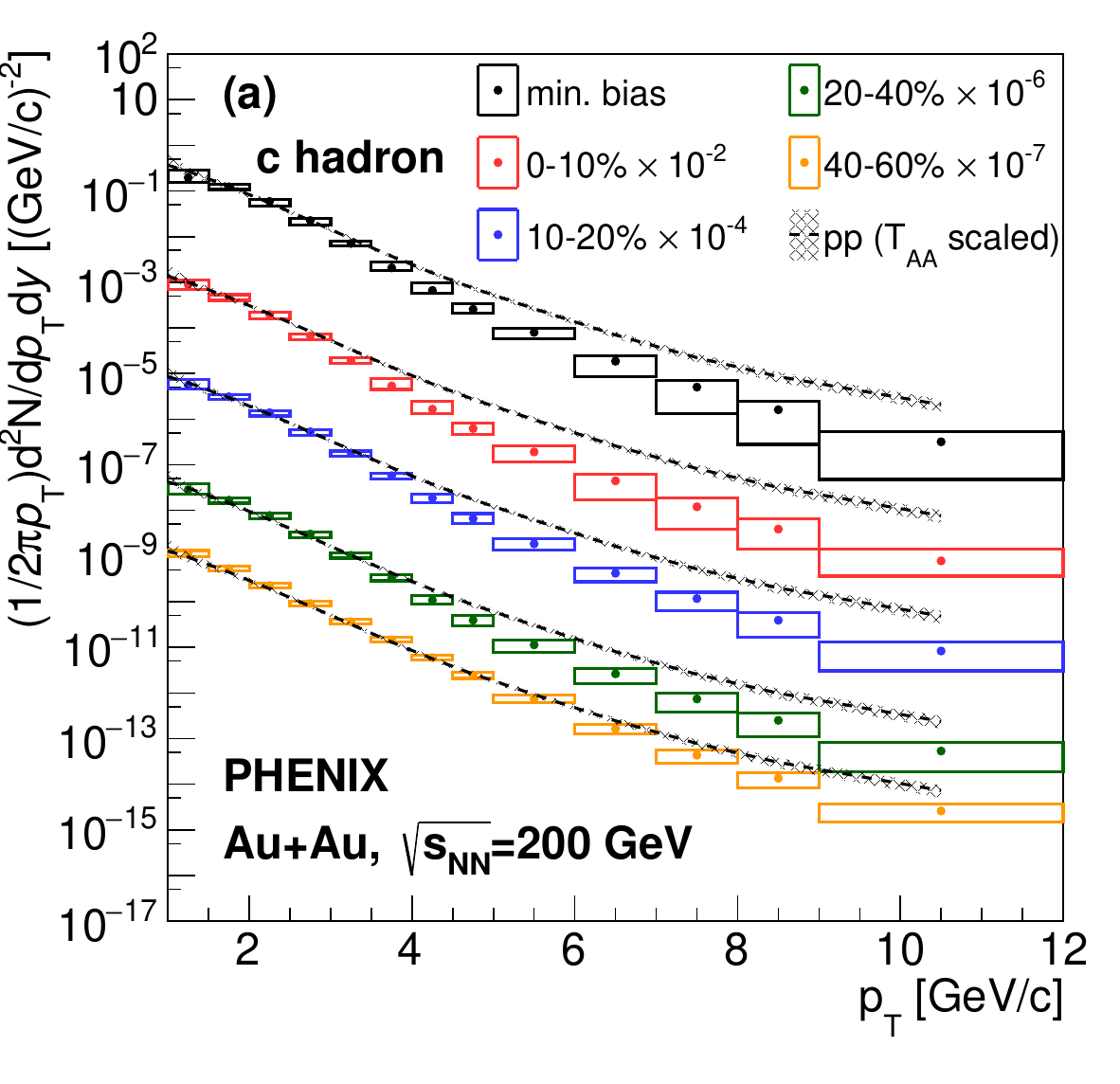}
\includegraphics[width=0.99\linewidth]{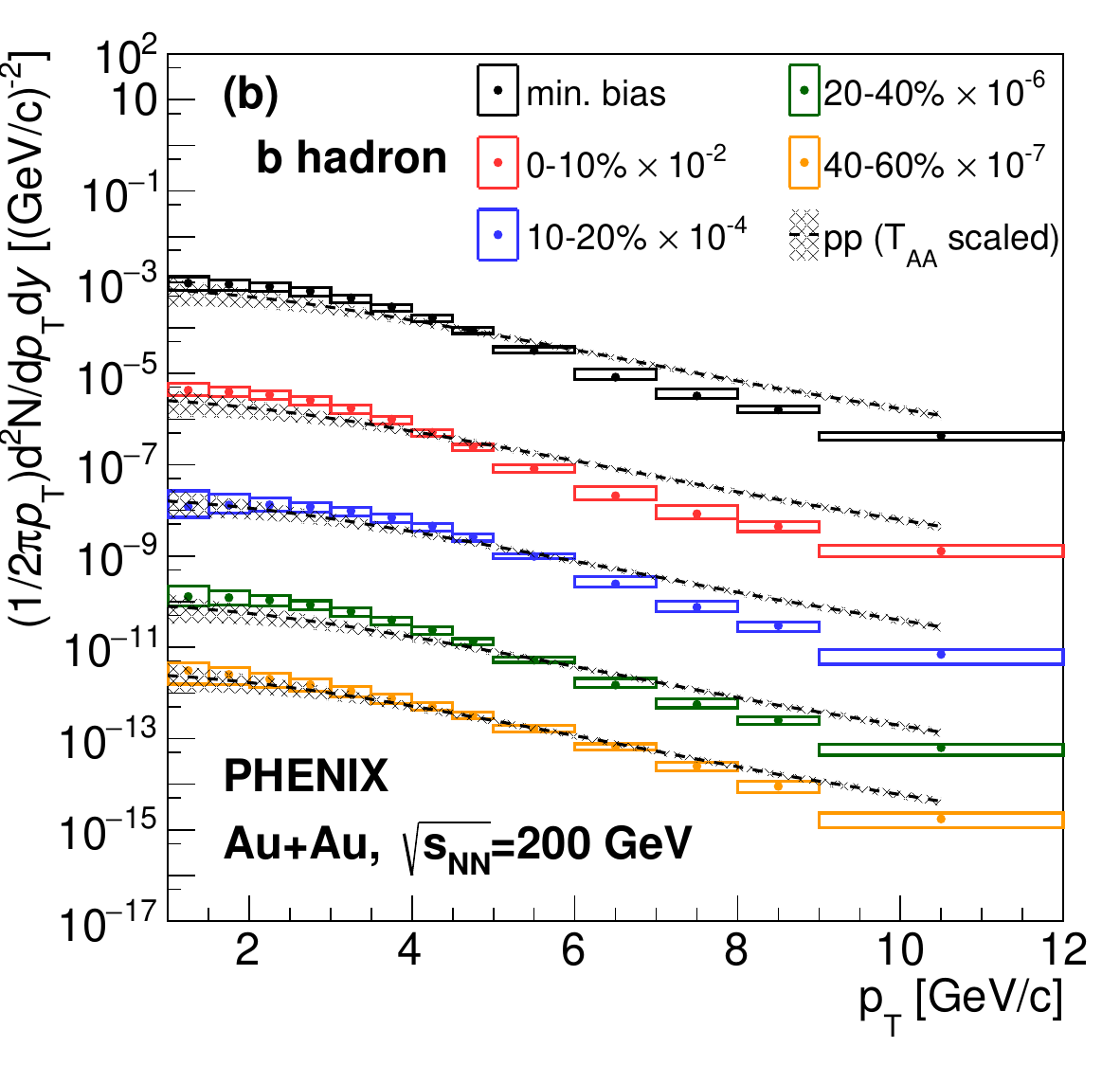}
\caption{Comparison of unfolded \pt spectrum of (a) charm hadrons and 
(b) bottom hadrons in \auau collisions to that scaled by $T_{AA}$ in \pp 
collisions~\cite{Aidala_2019}.
}
\label{fig:InvYield_chbh}
\end{minipage}
\end{figure*}

\begin{figure}[htb]
\hspace{0.4cm}
\centering
\includegraphics[width=1.0\linewidth]{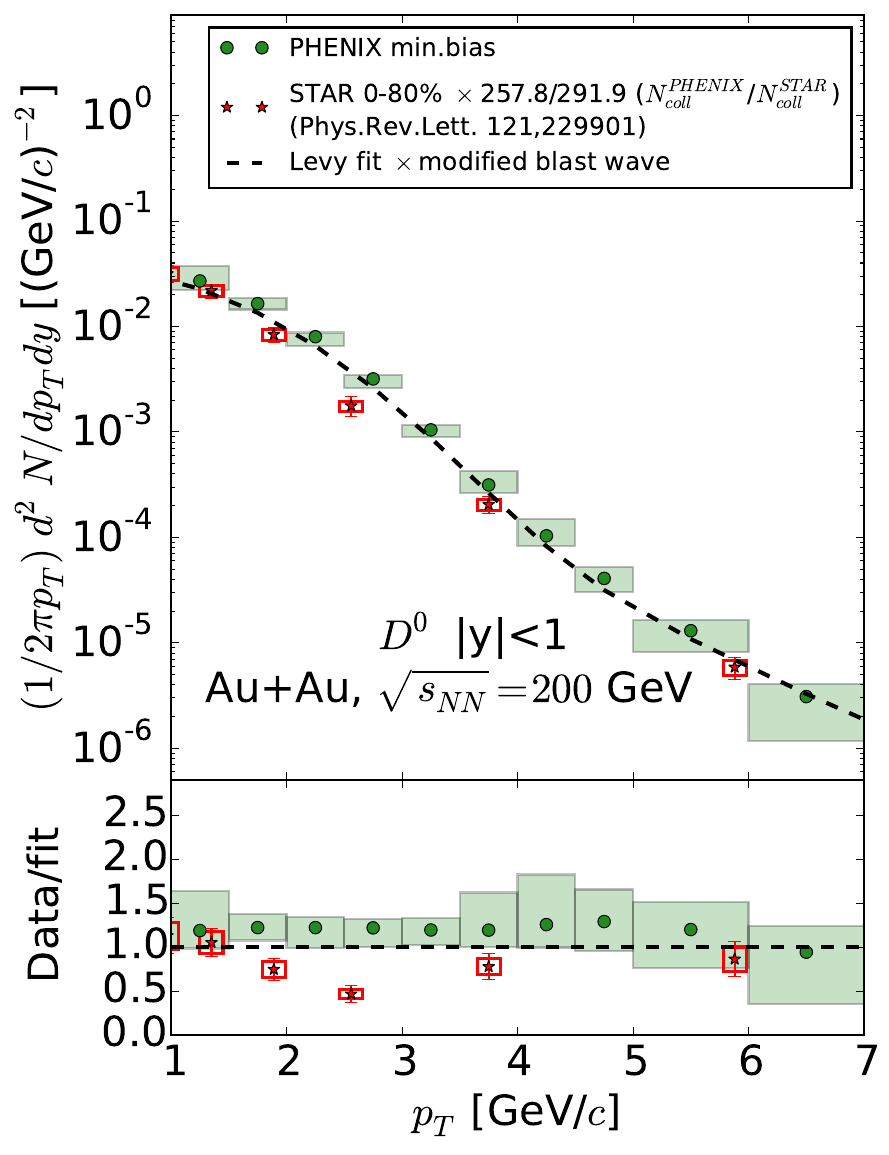}
\caption{Unfolded yield of $D^0$ mesons as a function of \pt 
at midrapidity $|y|<$ 1, compared to the measurement from 
STAR~\cite{Adamczyk:2014uip}.
}
\label{fig:compSTARD0}
\end{figure}

\begin{figure*}[htb]
\includegraphics[width=0.99\linewidth]{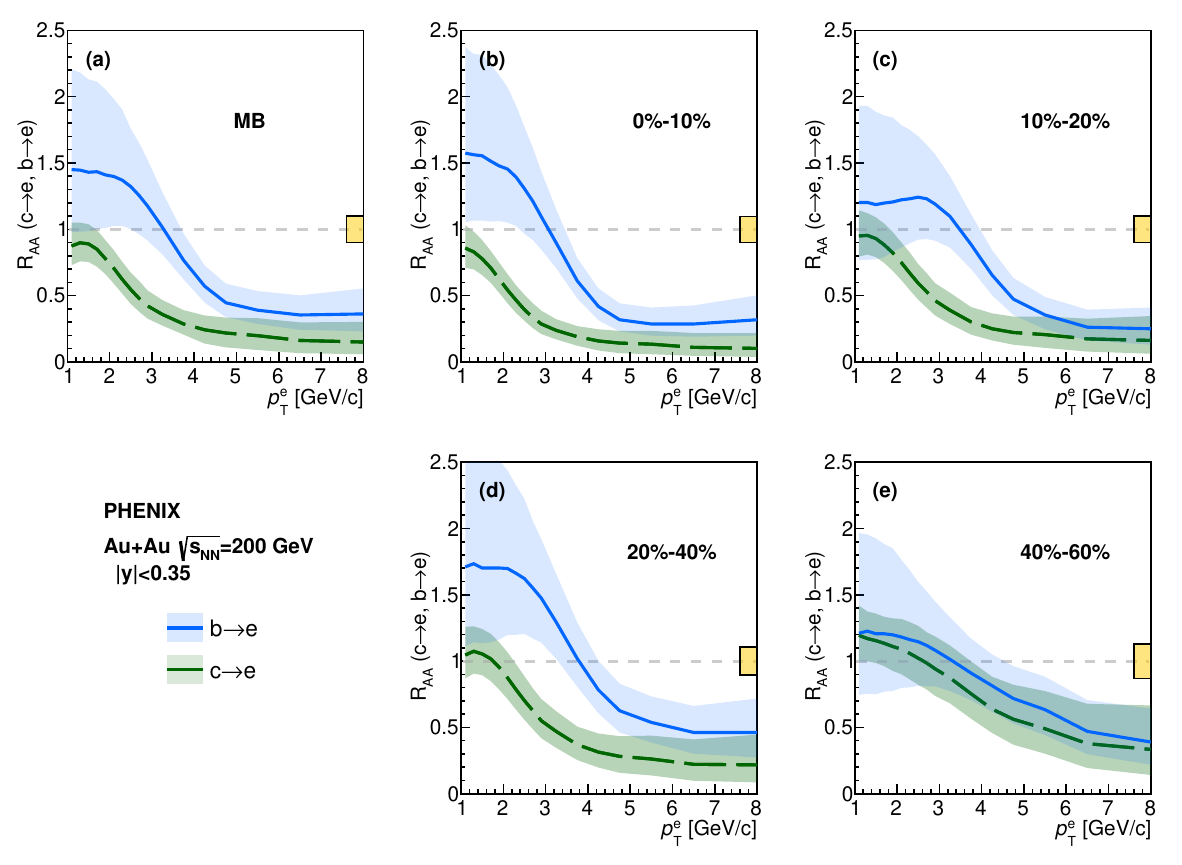}
\caption{The nuclear modification of \ctoe and \btoe as a function of 
\pt for different centrality classes. The yellow box at unity is the 
uncertainty on the total normalization.}
\label{fig:Raa_e_Pt}
\end{figure*}

\begin{figure}[htb]
\vspace{-0.3cm}
\includegraphics[width=1.0\linewidth]{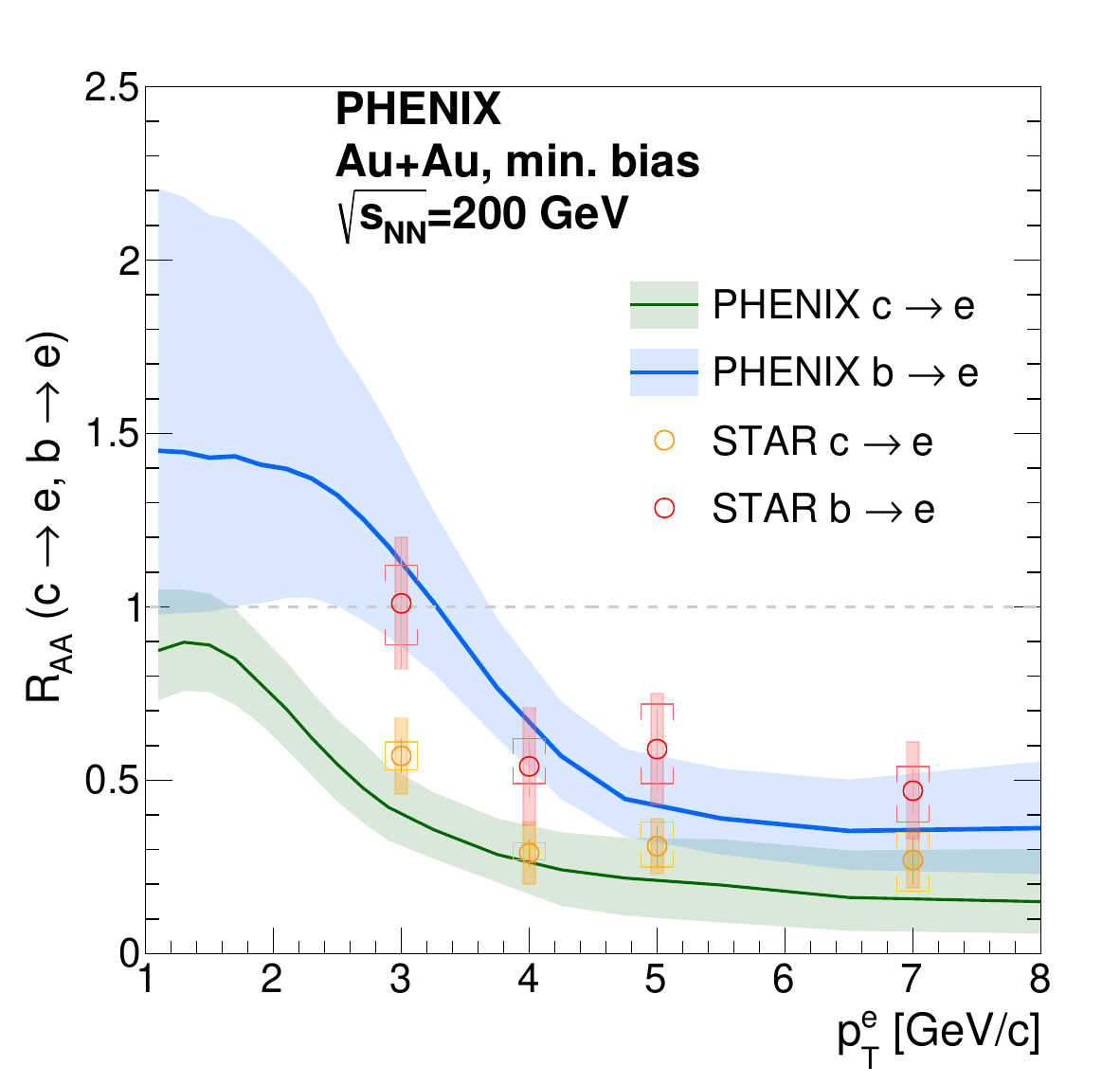}
\caption{The nuclear-modification factors of \ctoe and \btoe as a 
function of \pt in MB Au$+$Au collisions from this work 
compared with the corresponding measurement from the STAR 
collaboration~\cite{STAR_hfe_2021}.}
\label{fig:Raa_e_Pt_star}
\end{figure}

\begin{figure*}[hbt]
\begin{minipage}{0.99\linewidth}
\includegraphics[width=0.99\linewidth]{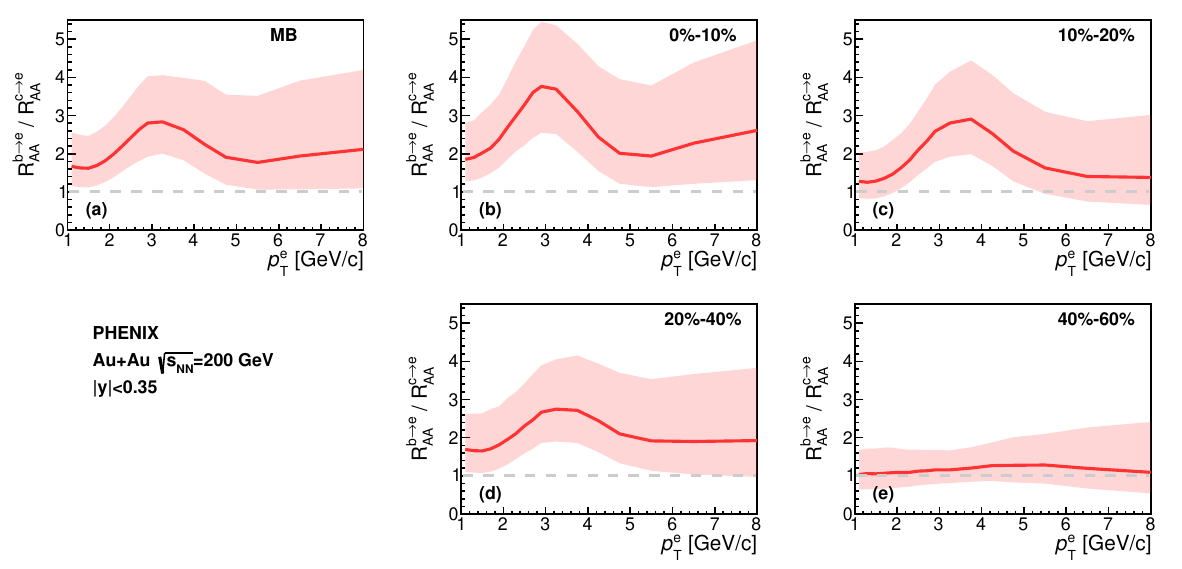}
\caption{\raa ratio of \btoe to \ctoe as a function of \pt for different 
centrality classes.
}
\label{fig:Raa_e_ratio}
\end{minipage}
\begin{minipage}{0.99\linewidth}
\includegraphics[width=0.99\linewidth]{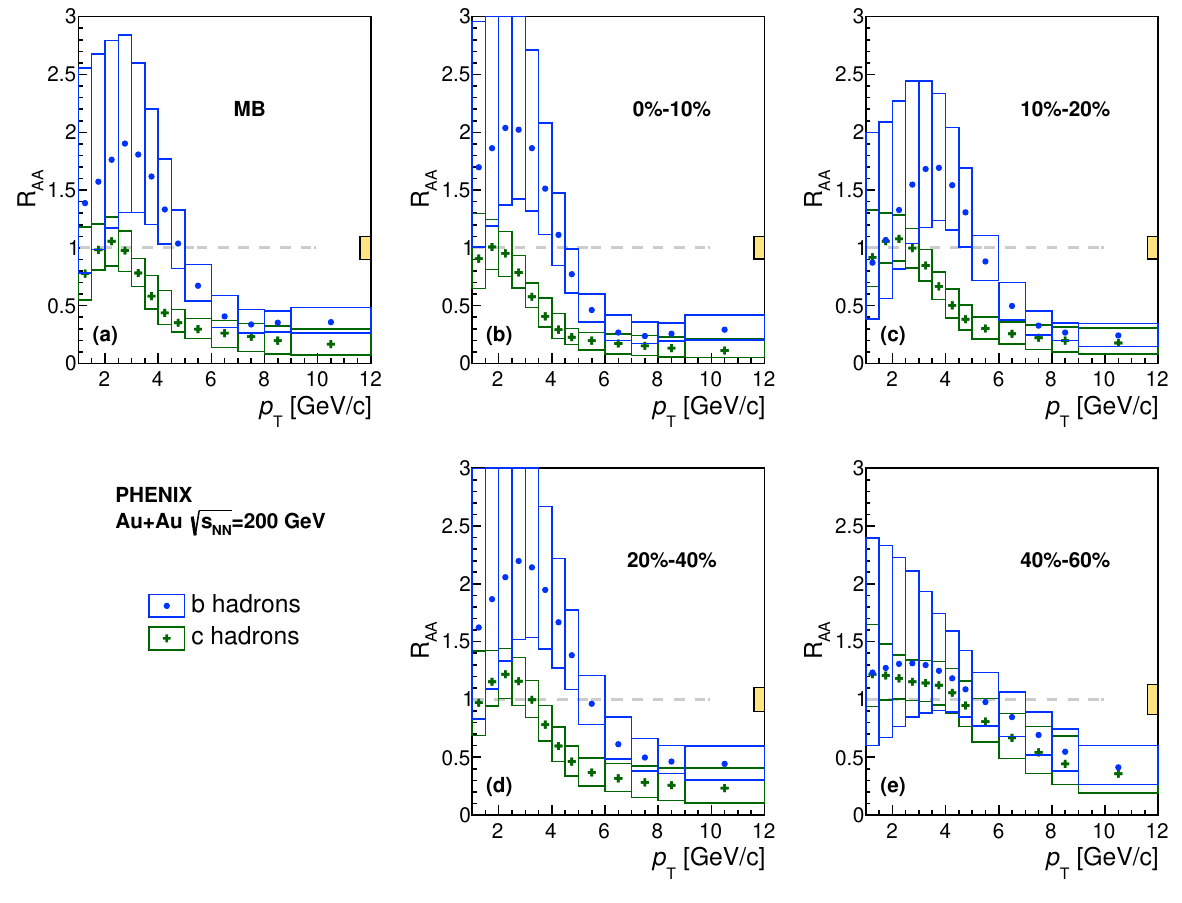}
\caption{The nuclear modification of charm and bottom hadrons as a 
function of \pt for different centrality classes. The yellow box at 
unity is the uncertainty on the total normalization.}
\label{fig:Raa_h_Pt}
\end{minipage}
\end{figure*}

\begin{figure*}[htb]
\includegraphics[width=0.99\linewidth]{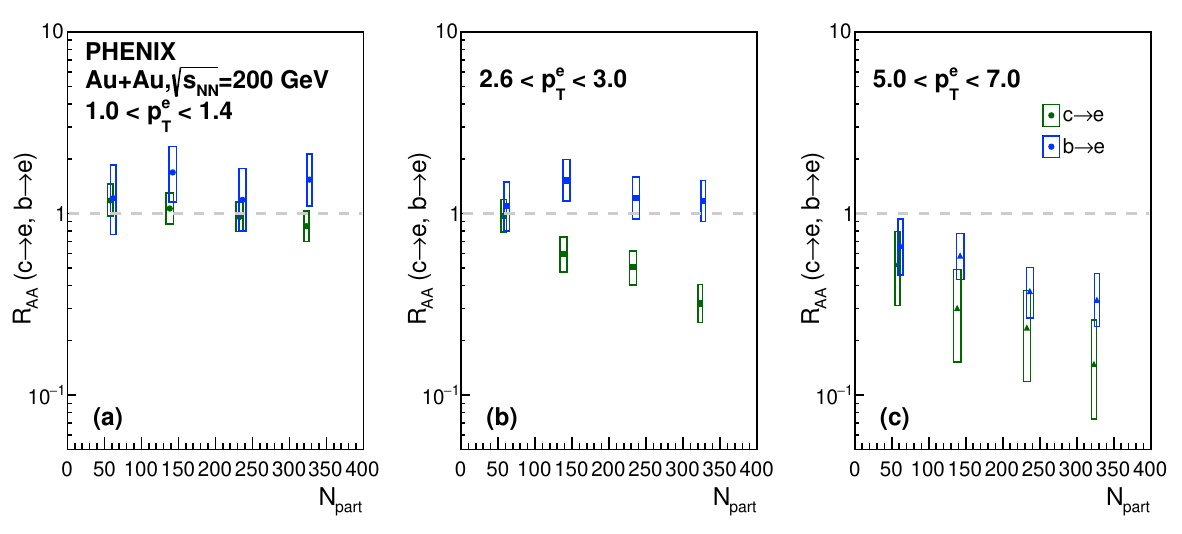}
\caption{The \raa for \ctoe and \btoe as a function of \npart in three 
different \pt ranges.  Data points for \ctoe and \btoe are shifted by -2 
and +2 from their respective \npart for clarity.
}
\label{fig:Raa_Npart}
\end{figure*}

\begin{figure}[hbt]
\includegraphics[width=1.0\linewidth]{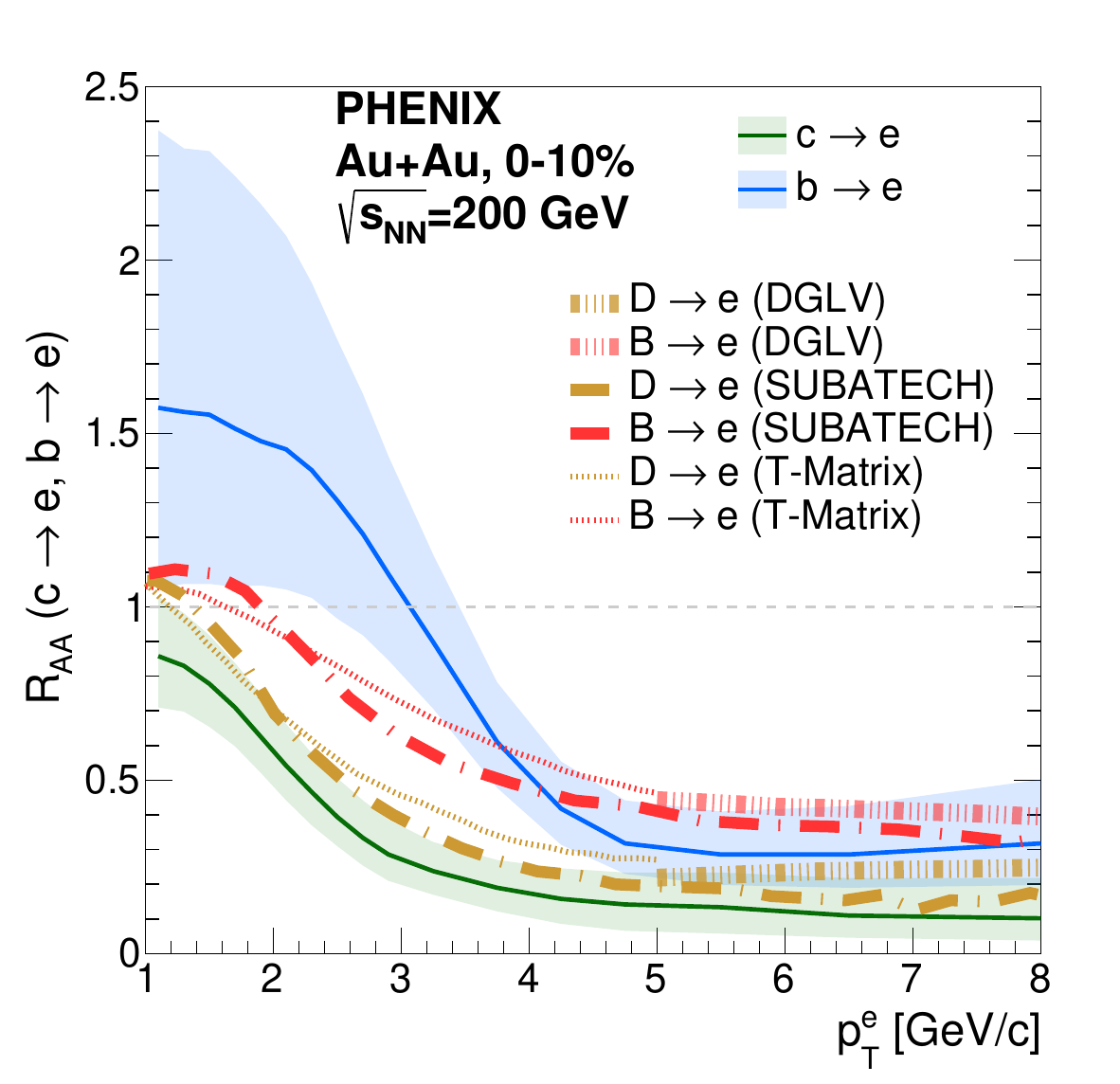}
\caption{Measured \raace and \raabe compared to 
theoretical-model calculations.}
\label{fig:RaaModel}
\end{figure}

The systematic uncertainties are independently evaluated for the 
measured data and the unfolding procedure. Figure~\ref{fig:SysErr} shows 
the contribution of each systematic uncertainty source. The total 
uncertainty is obtained by adding them in quadrature. Each source of 
uncertainty is discussed below.

\subsubsection{Background normalization}

Systematic uncertainties associated with modeling of the background 
processes are estimated from the difference between the nominal 
measurement and that obtained by repeating the unfolding procedure with 
systematic variation of the background \dcat normalization. The 
background \dcat template for each source of background is modified 
independently by $\pm$1$\sigma$ of the nominal value, and the unfolding 
procedure is repeated with the modified-background \dcat template. For 
each background source, the difference between the unfolding result 
using nominal-background templates and that with a modified-background 
template is taken as the systematic uncertainty. Estimates of background 
normalization uncertainty from all the background processes are added in 
quadrature to get a single value of the background normalization 
uncertainty.

\subsubsection{Measured yield of \cbtoe}

The unfolding procedure only considers statistical uncertainty on the 
measured yield of \cbtoe in the log-likelihood calculation. The 
systematic uncertainty on the measured yield of \cbtoe needs to be 
accounted for separately. To calculate the systematic uncertainty, an 
input \pt spectrum is modified by either kinking or tilting the 
spectrum. Tilting implies modifying the spectrum by pivoting the nominal 
spectrum about a given point such that the lowest \pt point goes up by 
the systematic uncertainty and the highest \pt point goes down by the 
same systematic uncertainty, while the intermediate points are modified 
with the linear interpolation of the two points. In contrast,
kinking implies that the modified spectrum is folded based on the 
nominal spectrum. The control point for both tilting and kinking is 
chosen at $\pt=1.8$~or~5.0~\gevc because analysis cuts are changed 
at these points. Once the spectra are modified with this tilting and 
kinking method, the unfolding procedure is run with 8 modified spectra, 
and the root mean square of the difference from the nominal result is 
assigned as a systematic uncertainty.

\subsubsection{Choice of prior}

In the Bayesian approach to unfolding, the prior is chosen to reflect 
a~priori knowledge of model parameters. In this analysis, {\sc 
pythia}-based distributions are used to model this initial knowledge. In 
theory, the optimal distributions obtained through the iterative 
unfolding procedure should be independent of the choice of the prior. 
However, residual model dependencies could be present. To account for 
any uncertainties due to the choice of the prior, the unfolding 
procedure is repeated with a modified prior, and the difference in the 
unfolded result from the nominal is assigned as a systematic 
uncertainty. The modified {\sc pythia} spectra are obtained by scaling 
heavy-flavor-hadron yields in {\sc pythia} with the blast-wave 
model~\cite{Adare:2013wca}.

\subsubsection{Regularization hyperparameter}

We control the strength of the regularization (spectrum smoothness) with 
a hyperparameter $\alpha$ of Eq.~(\ref{eq:10}).  The uncertainty due to 
$\alpha$ is determined by changing $\alpha$ by a half unit of the 
maximum-likelihood value which corresponds to 1$\sigma$ deviation. The 
differences of the unfolded results with these $\alpha$ values are taken 
as the systematic uncertainty of $\alpha$.


\section{Results}
\label{sec:result}

\subsection{Invariant yield}

The Bayesian unfolding is applied for MB \auau collisions as well as 
four centrality classes in \auau collisions. 
Figure~\ref{fig:InvYield_cebe} shows the invariant yields of electrons 
from charm and bottom hadron decays in \auau collisions at \snn = 
200~GeV. The line represents the median of the yield distribution 
at a given \pt and the band represents the 1$\sigma$ limits on the 
point-to-point correlated uncertainty. These yields are compared with 
the PHENIX \pp result scaled by the nuclear-overlap function, 
$T_{AA}$~\cite{Aidala_2019}. Both comparisons of the invariant yields of 
\ctoe and \btoe show substantial yield suppression at high \pt. The 
suppression increases at higher \pt and in more-central collisions.

The invariant yields of charm and bottom hadrons are unfolded 
point-by-point in 17 bins for each centrality class as shown in 
Fig.~\ref{fig:InvYield_chbh}. The point at each \pt bin is the most 
likely value of the hadron yields to describe the measured electron 
yields and \dcat distributions. Note that the hadron yields are 
integrated over all rapidity because the decay matrix used in the 
unfolding method handles all hadron rapidity decaying into electrons in 
the PHENIX acceptance.
 
Our unfolded charm-hadron yields have been compared with $D^0$ yields in 
Au$+$Au collisions measured by the STAR 
collaboration~\cite{Adamczyk:2014uip}. To compare them, {\sc pythia} is 
used to calculate the $D^0$ fraction within $|y|<$ 1 compared to all 
charm hadrons for the whole rapidity region. To match the centrality 
range, the STAR result is scaled by the ratio of the number of 
binary-collisions. This comparison is shown in 
Fig.~\ref{fig:compSTARD0}. For clarity, we have fit our unfolded $D^0$ 
yields with the modified Levy function used in 
Ref.~\cite{Adare:2015hla}. The ratio of the data to the fit is shown in 
the bottom panel of Fig.~\ref{fig:compSTARD0}. Within uncertainties, the 
unfolded $D^0$ yield is found to be in qualitative agreement with the 
$D^0$ yields~\cite{Adamczyk:2014uip}.

\subsection{Nuclear modification factor \raa vs. \pt}

To compare the yield suppression between charm and bottom quarks, the 
nuclear-modification factor \raa is calculated as
\begin{eqnarray}\label{eq:Raa}
 R_{AA}^{\ c{\rightarrow}e} &=& \frac{(1-F_{\rm AuAu})}{(1-F_{pp})} R_{AA}^{\rm HF}, \\
 R_{AA}^{\ b{\rightarrow}e} &=& \frac{F_{\rm AuAu}}{F_{pp}} R_{AA}^{\rm HF},
\end{eqnarray}
where $F_{\rm AuAu}$ ($F_{pp}$) is the bottom electron fraction in \auau 
(\pp), and $R_{AA}^{\rm HF}$ is the nuclear modification of inclusive 
heavy-flavor electrons (charm and bottom) whose yields are fully 
anticorrelated. The \raace and \raace are calculated by determining the full 
probability distribution assuming Gaussian uncertainty on 
$F_{\rm AuAu}$, $F_{pp}$, and $R_{AA}^{\rm HF}$. The median of the 
distribution is taken to be the center value with lower and upper 
one-$\sigma$ uncertainties of 16\% and 84\% of the distribution, 
respectively.

Figure~\ref{fig:Raa_e_Pt} shows \raace and \raabe as a function of \pt for 
MB \auau collisions as well as four centrality classes in \auau collisions. 
These results are improved by six times more Au$+$Au data than the previous 
analysis with a wider active area of the VTX detector~\cite{Adare:2015hla} 
and the latest $p$$+$$p$~\cite{Aidala_2019}. The $p$$+$$p$ reference was 
also improved by using the same VTX analysis technique with ten times more 
statistics than the previous $p$$+$$p$ result~\cite{Adare:2009ic}.

These results extend the \pt coverage down to 1~\gevc and the systematic 
bands are reduced by a factor of two. The systematic uncertainty of 
\raabe is large at low \pt because of the large uncertainty of $F_{pp}$ 
at low \pt, but the uncertainty of bottom electrons in Au$+$Au is 
independent of \pt. Significant suppression is seen for electrons from 
both charm and bottom decays at high \pt at MB and all centrality 
classes. The nuclear modification is consistent with unity within 
uncertainties at low \pt. Charm electrons show a stronger suppression 
than bottom electrons for $2<\pt<5$~\gevc in MB and 0\%--10\%, 
10\%--20\%, 20\%--40\% centrality classes, whereas charm and bottom 
suppression are similar at 40\%--60\%. Note that the prior information 
used in the unfolding is changed for these centralities. This change can 
possibly bias the center position of the resulting \ctoe and \btoe 
yields. If there is energy loss, then the \pt spectra are shifted to 
lower \pt. Therefore, the resulting \raa is suppressed at high \pt, but 
the yield is slightly enhanced at low \pt to conserve the total number 
of produced particles. For bottom hadrons, this enhancement can be seen 
at higher \pt than the charm hadrons due to the harder \pt slope.

The nuclear modification for charm and bottom electrons in 0\%--80\% 
Au$+$Au collisions was reported from the STAR 
collaboration~\cite{STAR_hfe_2021}. As Fig.~\ref{fig:Raa_e_Pt_star} 
shows, our unfolding results for charm and bottom electrons are in good 
agreement with the STAR measurements within uncertainties.

Figure~\ref{fig:Raa_e_ratio} shows the significance of the difference 
between \raace and \raabe, where the ratio of \raabe/\raace is calculated, 
leading to cancellation of the correlated uncertainty between \ctoe and 
\btoe yields. The data show that \raabe is at least one standard 
deviation higher than \raace in almost the entire \pt range for the most 
central events 0\%--40\%, with the largest difference at 3~\gevc.

To account for possible autocorrelations in the electron-decay 
kinematics, the \raa of parent charm and bottom hadrons are calculated 
with the unfolded yield of charm and bottom hadrons as shown in 
Fig.~\ref{fig:Raa_h_Pt}.  A significant difference of the yield 
suppression between charm and bottom hadrons is observed in the region 
$2<\pt<6$~\gevc in 0\%--40\% central collisions, similar to what is 
seen in the decay-electron space.


\subsection{Nuclear modification factor \raa vs. \npart}

The collision centrality is characterized by the number of nucleon 
participants in the collision (\npart) estimated using Monte-Carlo 
Glauber calculations. The \npart-dependent nuclear modifications \raace 
and \raabe are obtained in three \pt intervals as shown in 
Fig.~\ref{fig:Raa_Npart}.

In the low-\pt region (1.0--1.4~\gevc), there is no \npart dependence 
and no suppression for both \ctoe and \btoe, within uncertainties. The 
mid-\pt region (2.6--3.0~\gevc) shows a clear suppression of charm 
hadrons when the number of participants increases. The high-\pt region 
(5.0--7.0~\gevc) shows an increasing suppression of both charm and 
bottom hadrons with increasing collision centrality.

\subsection{Comparison to theoretical models}

Figure~\ref{fig:RaaModel} shows a comparison of data to three 
theoretical models: the T-Matrix approach, the SUBATECH model, and the 
DGLV model.  The T-Matrix approach is a calculation assuming formation 
of a hadronic resonance by a heavy quark in the QGP based on lattice 
quantum chromodynamics~\cite{vanhees_2008_tmat}. The SUBATECH model 
employs a hard thermal loop calculation for the collisional energy 
loss~\cite{Gossiaux:2008jv}. The DGLV model calculates both the 
collisional and radiative energy loss assuming an effectively static 
medium~\cite{PhysRevC.90.034910}. Because the DGLV model includes only  
energy loss and does not include the back reaction in the medium, the 
curves are only shown for $p_T>5$~\gevc. All models expect a quark mass 
ordering for the energy loss in the QGP medium, as observed in the data.  
The SUBATECH and DGLV calculations for charm suppression agree with the 
data. The T-Matrix approach is slightly higher than the data for \pt $>$ 
3~\gevc. The measured bottom nuclear modification is larger than the 
calculations at $p_T<4$~\gevc, although the uncertainty in the 
measurement is large for $p_T<2$~\gevc.

\section{Summary and Conclusions}
\label{sec:summary}

This article reported the results of measurements of the separated 
invariant yields and nuclear-modification factors of charm and bottom 
hadron-decay electrons in \auau collisions at \sqsn = 200 GeV at 
midrapidity. The measurements were performed by the use of a Bayesian 
unfolding method to extract the invariant yield of parent charm and 
bottom hadrons from \pt and transverse distance of the closest approach 
\dcat distributions of decay electrons.

The nuclear-modification factors \raa have been calculated from the 
invariant yield in \auau and the $T_{AA}$ scaled yield in \pp. The 
comparison between \raace and \raabe indicates that charm hadrons are 
more suppressed than bottom hadrons by at least one standard deviation 
for 0\%--40\% central collisions. Quark-mass ordering of suppression is 
also seen in the \raa of the parent charm and bottom hadrons, where 
there is a pattern of \raa consistent with unity for $p_T<1.4$~\gevc for 
both charm and bottom, charm suppression for $2.6<p_T<3.0$~\gevc, and 
suppression of both charm and bottom for $p_T>5.0$~\gevc. These results 
suggest that charm quarks lose more energy than bottom quarks when 
crossing the hot and dense medium created in 200~GeV \auau collisions in the 
intermediate-\pt region.  The theoretical models used to compare with 
our data are based on different energy-loss mechanisms and all agree 
with the mass ordering and the charm suppression for the entire \pt 
range covered by this measurement. However, the same models overestimate 
the bottom-quark suppression in the intermediate \pt region.


\begin{acknowledgments}

We thank the staff of the Collider-Accelerator and Physics
Departments at Brookhaven National Laboratory and the staff of
the other PHENIX participating institutions for their vital
contributions.  
We acknowledge support from the Office of Nuclear Physics in the
Office of Science of the Department of Energy,
the National Science Foundation,
Abilene Christian University Research Council,
Research Foundation of SUNY, and
Dean of the College of Arts and Sciences, Vanderbilt University
(USA),
Ministry of Education, Culture, Sports, Science, and Technology
and the Japan Society for the Promotion of Science (Japan),
Natural Science Foundation of China (People's Republic of China),
Croatian Science Foundation and
Ministry of Science and Education (Croatia),
Ministry of Education, Youth and Sports (Czech Republic),
Centre National de la Recherche Scientifique, Commissariat
{\`a} l'{\'E}nergie Atomique, and Institut National de Physique
Nucl{\'e}aire et de Physique des Particules (France),
J. Bolyai Research Scholarship, EFOP, the New National Excellence
Program ({\'U}NKP), NKFIH, and OTKA (Hungary),
Department of Atomic Energy and Department of Science and Technology
(India),
Israel Science Foundation (Israel),
Basic Science Research and SRC(CENuM) Programs through NRF
funded by the Ministry of Education and the Ministry of
Science and ICT (Korea),
Ministry of Education and Science, Russian Academy of Sciences,
Federal Agency of Atomic Energy (Russia),
VR and Wallenberg Foundation (Sweden),
University of Zambia, the Government of the Republic of Zambia (Zambia),
the U.S. Civilian Research and Development Foundation for the
Independent States of the Former Soviet Union,
the Hungarian American Enterprise Scholarship Fund,
the US-Hungarian Fulbright Foundation,
and the US-Israel Binational Science Foundation.

\end{acknowledgments}



%
 
\end{document}